\pdfoutput=1
\documentclass[journal]{IEEEtran}
\usepackage[utf8]{inputenc}
\usepackage{amsmath}
\usepackage{amsfonts}
\usepackage{bbding}
\usepackage{amssymb}
\usepackage{array}
\usepackage{multirow}
\usepackage{graphicx}
\usepackage{algorithm}
\usepackage{algorithmic}
\usepackage{color}
\allowdisplaybreaks

\usepackage{amsthm}
\usepackage{stfloats}
\usepackage{subcaption}

\begin{document}
\title{
Knowledge-Assisted Deep Reinforcement Learning in 5G Scheduler Design: From Theoretical Framework to Implementation}

\author{
    \IEEEauthorblockN{Zhouyou Gu, Changyang She, Wibowo Hardjawana, Simon Lumb, David McKechnie, \\Todd Essery, Branka Vucetic}\\
    \thanks{This work is part of a project directly funded by Telstra Corporation Ltd., titled "Development of an Open Programmable Scheduler for LTE Networks". It is supported by an Australian Government Research Training Program Scholarship and two Postgraduate Research Supplementary Scholarships of The University of Sydney. It is also supported by an Australian Research Council Discovery Early Career Research Award (DE150101704) and partially supported by an Australian Research Council Laureate Fellowship (FL160100032). \emph{(Corresponding author: Changyang She.)}}
    \thanks{Z. Gu, C. She, W. Hardjawana, and B. Vucetic are with the School of Electrical and Information Engineering, University of Sydney, Sydney, NSW 2006, Australia (email: \{zhouyou.gu, changyang.she, wibowo.hardjawana, branka.vucetic\}@sydney.edu.au).}
    \thanks{S. Lumb, D. McKechnie, T. Essery are with Telstra Corporation Ltd, Australia (email: \{simon.lumb, david.mckechnie, todd.essery\}@team.telstra.com).}
    \thanks{Source codes are available at {github.com/zhouyou-gu/drl-5g-scheduler}}
}
\maketitle
\begin{abstract}
In this paper, we develop a knowledge-assisted deep reinforcement learning (DRL) algorithm to design wireless schedulers in the fifth-generation (5G) cellular networks with time-sensitive traffic. Since the scheduling policy is a deterministic mapping from channel and queue states to scheduling actions, it can be optimized by using deep deterministic policy gradient (DDPG). We show that a straightforward implementation of DDPG converges slowly, has a poor quality-of-service (QoS) performance, and cannot be implemented in real-world 5G systems, which are non-stationary in general. To address these issues, we propose a theoretical DRL framework, where theoretical models from wireless communications are used to formulate a Markov decision process in DRL. To reduce the convergence time and improve the QoS of each user, we design a knowledge-assisted DDPG (K-DDPG) that exploits expert knowledge of the scheduler design problem, such as the knowledge of the QoS, the target scheduling policy, and the importance of each training sample, determined by the approximation error of the value function and the number of packet losses. Furthermore, we develop an architecture for online training and inference, where K-DDPG initializes the scheduler off-line and then fine-tunes the scheduler online to handle the mismatch between off-line simulations and non-stationary real-world systems. Simulation results show that our approach reduces the convergence time of DDPG significantly and achieves better QoS than existing schedulers (reducing $30\% \sim 50\%$ packet losses). Experimental results show that with off-line initialization, our approach achieves better initial QoS than random initialization and the online fine-tuning converges in few minutes.
\end{abstract}

\section{{Introduction}}\label{sec:introduction}
The 5th generation (5G) cellular networks are expected to support emerging applications with time-sensitive traffic, such as autonomous vehicles, factory automation, tactile internet, and virtual/augmented reality \cite{aijaz2018tactile,bennis2018ultrareliable,she2017radio}. Time-sensitive traffic has stringent quality-of-service (QoS) requirements, including delay, reliability, and jitter \cite{3gpp.22.104}, which are different from the design goal of the previous generations of cellular networks, i.e., pursuing higher data rates. Existing schedulers like proportional fair \cite{huang2018gpf}, round-robin \cite{kawser2012performance}, earliest-deadline-first \cite{andrews2000probabilistic} and maximum throughput \cite{schwarz2010low} were not developed for time-sensitive traffic. Thus, wireless schedulers should be re-designed to meet the QoS requirements of time-sensitive traffic in 5G.

A wireless scheduler is a multi-dimensional function that takes the queue state information (QSI) and the channel state information (CSI) as its input and outputs the amount of resources allocated to users. Such a problem can be formulated as an optimal control problem of a Markov decision process (MDP), which can be solved by reinforcement learning \cite{she2020deep}. Classic reinforcement learning algorithms and dynamic programming suffer from the curse of dimensionality, and are only applicable to problems with small state-action spaces. To overcome this difficulty, one can apply deep Q-learning, where the state-action value function is approximated by a neural network (NN) \cite{mnih2015human}. With deep Q-learning, the scheduler needs to find the optimal action that maximizes the value function in each transmission time interval (TTI) in 5G New Radio (NR). Since the action space could be very large, the scheduler can hardly solve the optimization problem in one TTI. More recently, actor-critic deep reinforcement learning (DRL) algorithms have been developed to handle the above issue, where the policy and the long-term reward are approximated by two NNs, respectively \cite{sutton2011reinforcement}. If the optimal policy is deterministic, which is the usual case in most optimal control problems \cite{sutton2011reinforcement}, the actor-critic DRL algorithms become the deep deterministic policy gradient (DDPG) algorithm \cite{lillicrap2015continuous}.

Since DDPG is a model-free algorithm that does not require the transition probabilities of the MDP, it usually converges very slowly. However, communication environments in real-world networks are non-stationary. Once communication environments change, the pre-trained scheduling policy cannot achieve good QoS. Therefore, to apply DDPG in scheduler design, we need to fine-tune the policy online and reduce its convergence time in the real-world networks. On the other hand, the feed-forward inference of the scheduler, i.e., computing the output of the NN for a given input, at the base station (BS) must be completed within each TTI ($0.125 \sim 1$~ms). This poses a new challenge to implement learning-based schedulers in real-world 5G systems \cite{huang2018gpf}.

\subsection{Related Works}\label{sec:related_works}

\subsubsection{Schedulers for Time-sensitive Traffic} How to develop scheduler for time-sensitive traffic in wired networks has been discussed in time-sensitive networking standardization \cite{nasrallah2018ultra,specht2017synthesis}. For example, the authors of \cite{specht2017synthesis} developed an urgent-based scheduler and analyzed the worst-case latency. To serve time-sensitive traffic with deterministic packet arrival processes in wireless communications, wireless schedulers were investigated in \cite{khoshnevisan20195g,ginthor2019analysis}. Specifically, the semi-persistent scheduling has been adopted in 5G NR for periodic transmissions of control signals \cite{khoshnevisan20195g}. A system-level simulator for evaluating the end-to-end (E2E) performance of schedulers when supporting deterministic traffic was carried out in \cite{ginthor2019analysis}. In the above publications, either the channels (wired networks in \cite{nasrallah2018ultra,specht2017synthesis}) or the arrival processes (control signaling and data packets in  \cite{khoshnevisan20195g,ginthor2019analysis}) are assumed to be deterministic. However, both wireless channels and arrival processes are stochastic in various 5G applications, such as machine-type communications, vehicle safety applications, and ultra-reliable and low-latency communications \cite{3GPP2012MTC,Hassan2013A,3GPP2017Scenarios}. For these applications, how to achieve low latency and low jitter with high reliability remains an open challenge.

\subsubsection{DDPG in Wireless Networks}
DDPG has been widely applied to solve optimal control problems in wireless networks. The authors of \cite{tseng2019radio} used DDPG to select a radio resource scheduling policy from existing schedulers and resource allocation policies.
Considering that network slicing will be adopted in 5G to serve different kinds of services, DDPG was used to allocate resources among different slices in \cite{qi2019deep} and among different users in \cite{li2020deep}.
To further implement DDPG in wireless networks, an online architecture was implemented in virtual radio access networks for jointly controlling computing resources and the modulation and coding scheme, where a controller sends actions to virtual BSs every $20$~seconds \cite{ayala2019vrain}. Another online DDPG architecture for network slicing was implemented in the 4G radio access networks in \cite{foukas2019iris}, where the controller allocates resources among different slices every $20 \sim 100$~ms. None of existing architectures can be used for 5G scheduler design, where the scheduler takes actions in every TTI at a time resolution of $\sim0.1$~ms. Thus, an online architecture that enables DDPG in 5G scheduler design is much needed.
\begin{table*}[ht]
\vspace{0.2cm}\small
\caption{Issues of Straightforward Implementation of DDPG and Solutions to Address Them}
\label{tab:structure}
\begin{minipage}{\textwidth}
\begin{center}
\begin{tabular}{|p{7.8cm}|p{7.5cm}|}
\hline
Issues & Our Solutions 
\\\hline
Section \ref{subsec:issue_problem_formulation} Issues in Problem Formulation & Section \ref{sec:framework} Theoretical DRL Framework  
\\\hline
$\;$1) Large action space & $\;$1) Action space reduction \\\hline
$\;$2) Low flexibility of states & $\;$2) Generalization of states \\\hline
$\;$3) Poor reliability evaluation & $\;$3) Theoretical formulation for reliability evaluation \\\hline
Section \ref{subsec:issue_training} Issues in Training Algorithm & Section \ref{sec:k-ddpg} Knowledge-assisted DDPG  \\\hline
$\;$1) Unaware of individual QoS & $\;$1) Multi-head critic for individual QoS evaluation \\\hline
$\;$2) Delayed reward/Sparse reward & $\;$2) Reward shaping for instantaneous non-zero feedback\\\hline
$\;$3) Inaccurate critic at rarely visited state-action pairs & $\;$3) Importance sampling \\\hline
Section \ref{subsec:issue_online_implementation} Issues in Online Implementation & Section \ref{sec:oneline_archiecture} Online Architecture  \\\hline
$\;$1) Poor initial QoS & $\;$1) off-line initialization \& online fine-tuning \\\hline
$\;$2) Long processing time in each TTI & $\;$2) Parallel processing in the BS \\\hline
\end{tabular}
\vspace{-0.2cm}
\end{center}
\end{minipage}
\end{table*}

\subsubsection{Knowledge-Assisted Learning in Communications}
How to improve the training efficiency with the assistance of expert knowledge in vertical industries remains an important issue~\cite{ng1999policy}. The knowledge is referred to as the design principles and insights that are exploited by human experts to design learning algorithms, rather than specific data or information. Based on the expert knowledge of physical layer communications, the authors in  \cite{he2019model,he2020model} designed the structures of NNs to improve the training efficiency. Since the experts in wireless communications have developed a lot of optimization algorithms and heuristic solutions, the existing policies can be used to generate training samples for DRL algorithms \cite{gu2019intelligent}. Nevertheless, how to establish a DRL framework for scheduler design, and how to improve the training efficiency and the QoS of each user by exploiting expert knowledge in wireless communications require further research.

\subsection{Our Contributions}
In this paper, we first investigate a straightforward implementation of DDPG in 5G scheduler design for time-sensitive traffic, where no communication model or expert knowledge of the wireless scheduler design problem is exploited. The straightforward implementation suffers from several issues related to problem formulation, training, and online implementation, which are listed along with our solutions to them in Table \ref{tab:structure}. The main contributions of this paper are summarized as follows.
\begin{itemize}
\item We establish a theoretical DRL (T-DRL) framework for wireless scheduler design with time-sensitive traffic in 5G systems, where existing theoretical models and results in wireless communications are used to formulate the optimal control problem. Based on the models and formulations, we prove that the problem is Markovian, and hence we can apply DRL algorithms to solve it \cite{sutton2011reinforcement}. Different from the existing model-based dynamic programming, where the ``model" is described by the transition probabilities of the MDP \cite{sutton2011reinforcement}, T-DRL does not need the transition probabilities, but uses well-known channel and queueing models.
\item To improve the QoS of users and to reduce the convergence time, we design a knowledge-assisted DDPG (K-DDPG) algorithm that integrates DDPG with expert knowledge such as the knowledge of the QoS of each user, the target scheduling policy, and the importance of training samples (determined by the approximation error of the critic and the number of packet losses). To achieve this, multi-head critic, reward shaping and importance sampling are used in {K-DDPG}.
\item To fine-tune the scheduler in real-world networks, we develop an architecture that enables online training and inference of K-DDPG. An edge server in the architecture first initializes the scheduler off-line in a simulation platform built upon the T-DRL framework. Then, it keeps fine-tuning the scheduler according to feedback from real-world networks. Meanwhile, the BS executes the scheduling policy at every TTI
and shares the feedback from real-world networks to the edge server.
\item We build a prototype of the proposed architecture based on a standard-compliant cellular network software suite that is able to communicate with commercial devices~\cite{gomez2016srslte}. In the prototype, the online training converges in few minutes and the online inference can be executed in each TTI. Thus, our approach can be applied to scheduler design in 5G NR.
\end{itemize}

\section{Scheduler for Time-Sensitive Traffic in 5G}
\subsection{Wireless Scheduler in 5G NR}
\begin{figure}[!hbt]
\centering
\includegraphics[scale=0.8]{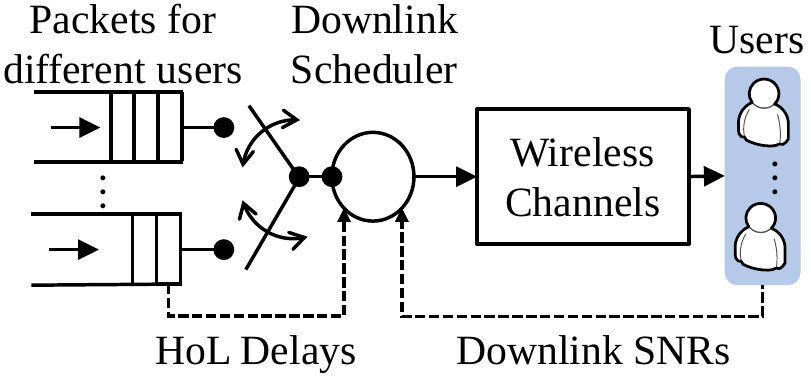}
\caption{Illustration of downlink scheduler.}
\label{fig:scheduler}
\vspace{-0.2cm}
\end{figure}
We consider a downlink scheduler in 5G NR, where $K$ users are served by one BS, as shown in Fig.~\ref{fig:scheduler}.
Packets of the $k$-th user are waiting in the $k$-th queue in the buffer of the BS, and each queue is served according to the first-in-first-out (FIFO) order.
The duration of one slot is equal to the duration of one TTI in 5G NR, and is denoted by $\Delta^t$. 

We leverage indicators $x_{k}(t), k=1,...,K$, to represent whether users are scheduled in the $t$-th slot.
If the $k$-th user is not scheduled in the $t$-th slot, $x_{k}(t)=0$. Otherwise, $x_{k}(t)=1$ and one packet will be transmitted to the user.
The packet size of the $k$-th user and the number of resource blocks (RBs) allocated to it in the $t$-th slot are denoted by $L_k$ (bits) and $n_k(t)$, respectively. 
Since orthogonal frequency division multiplexing is adopted in 5G NR systems, $n_k(t)$ can be adjusted in each slot by subcarrier allocation. As illustrated in Fig.~\ref{fig:scheduler}, a scheduler determines $x_{k}(t)$ and $n_k(t)$ according to QSI and CSI of all users, such as head-of-line (HoL) delays and downlink signal-to-noise ratios (SNRs) of users at the $t$-th slot, denoted by $d_k(t)$ and $\phi_k(t)$, respectively, $k=1,\dots,K$.

\subsection{QoS Requirements of Time-Sensitive Traffic}\label{subsec:qos_tsn}
The time-sensitive traffic has stringent QoS requirements, including delay, jitter, and reliability \cite{3gpp.22.104,nasrallah2018ultra,neumann2018towards}. To satisfy the delay requirement, the delay experienced by packets should be larger than a minimum delay bound and smaller than a maximum delay bound, which are denoted by $D_\text{min}$ and $D_\text{max}$, respectively \cite{3gpp.22.104,neumann2018towards}.
Such a constraint also guarantees that the jitter does not exceed $D_{\max} - D_{\min}$. The reliability of a user is defined as the packet loss probability. A packet is lost if $d_k(t) \notin  [D_{\min}, D_{\max}]$ or the decoding at the receiver fails.
For typical time-sensitive traffics, as shown in 3GPP standards~\cite{3gpp.22.104}, the maximum delay bound is around $1$ to $10$~ms, the jitter needs to be less than two TTIs and the target reliability varies from $99.9\%$ to $99.999\%$.

\section{Straightforward Implementation of DDPG for Scheduler Design}\label{sec:model_free_problem}
In this section, we introduce a straightforward implementation of DDPG for scheduler design.

\subsection{Preliminaries of DDPG}
\begin{figure}[!htb]
\centering
\includegraphics[scale=0.8]{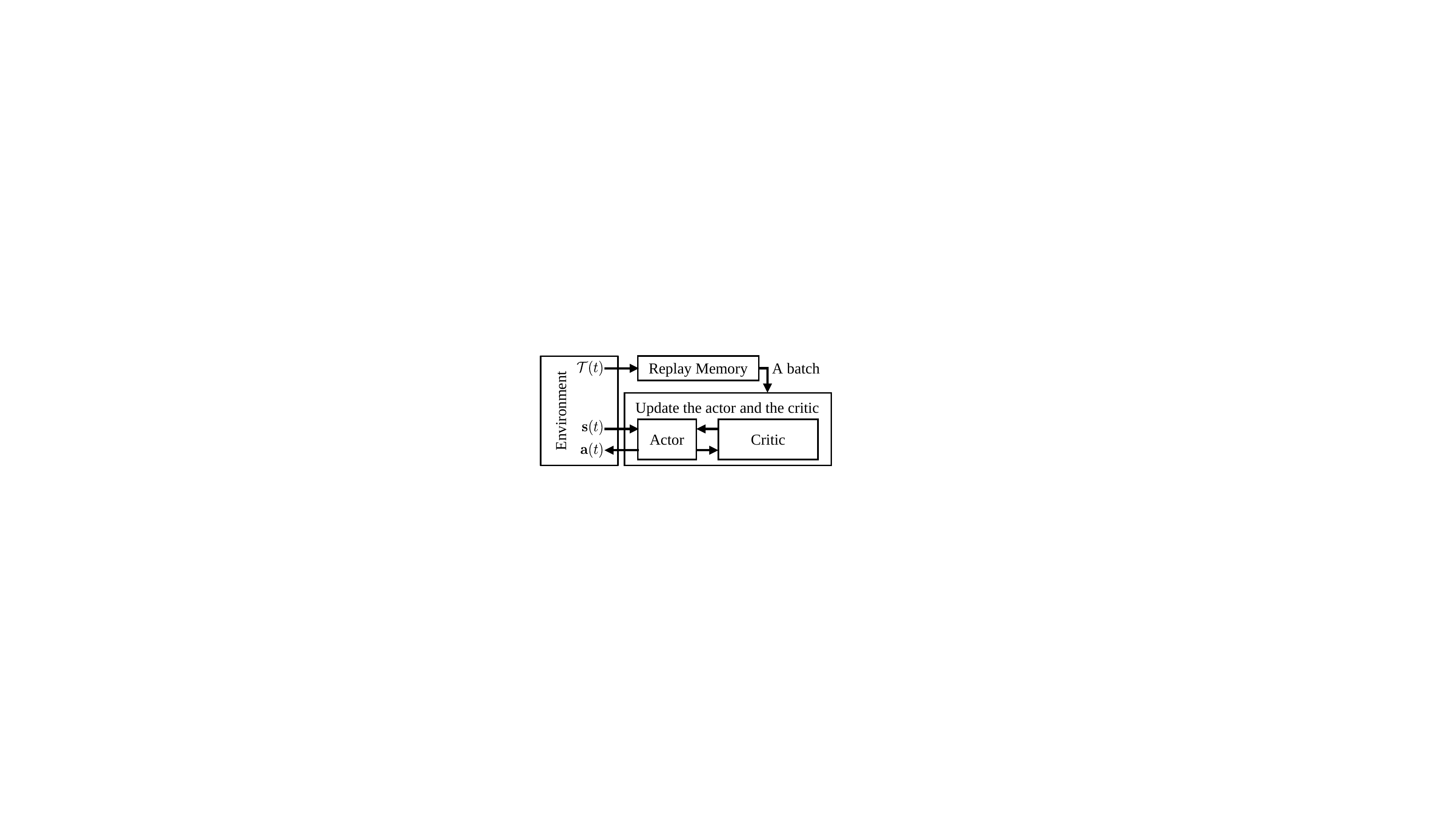}
\caption{Illustration of DDPG.}
\label{fig:ddpg}
\vspace{-0.2cm}
\end{figure}
\subsubsection{Training of DDPG} As shown in Fig.~\ref{fig:ddpg}, DDPG is an actor-critic reinforcement learning algorithm \cite{lillicrap2015continuous}, where the actor and the critic are two NNs that determine the action in a given state and evaluate the long-term reward of the state-action pair (taking an action in a state), respectively.
Given a system state, the action to be executed is obtained from the actor,
\begin{equation} \label{eq:actor}
\begin{aligned}
{\bf{a}}(t) = \mu\big({\bf{s}}(t)|\theta^\mu\big) \ ,
\end{aligned}
\end{equation}
where ${\bf{s}}(t)$ and ${\bf{a}}(t)$ are the state and action in the $t$-th slot, and $\mu(\cdot|\theta^\mu)$ represents the actor and $\theta^\mu$ represents the parameters of the actor, i.e., weights and biases.

Given ${\bf{s}}(t)$ and ${\bf{a}}(t)$, the long-term reward is estimated by a state-action value function,
\begin{equation}\label{eq:q_pi}
\begin{aligned}
&Q\big({\bf{s}}(t),{\bf{a}}(t)\big)=\\
&\mathop{\mathbb{E}} \Big[ \sum_{i=0}^{\infty}\gamma^i r(t+i)\ \Big|\  {\bf{s}}(t), {\bf{a}}(t), {\bf{a}}(t') = \mu\big({\bf{s}}(t')|\theta^{\mu}\big), \forall t'>t\Big],
\end{aligned}
\end{equation}
where $r(t+i)$ is the instantaneous reward in the $(t+i)$-th slot and $\gamma$ is the discount factor that measures the importance of the future rewards.
The state-action value function in \eqref{eq:q_pi} is approximated by the critic, $Q(\cdot|\theta^{Q})$, where $\theta^{Q}$ represents the parameters of the critic.

DDPG initializes the parameters of the NNs, $\theta^\mu$ and $\theta^Q$, as random values.
In each time slot, the system observes the current state and generates an action according to ${\bf{a}}(t)=\mu({\bf{s}}(t)|\theta^{\mu})+\mathcal{N}(t)$, where $\mathcal{N}(t)$ is a exploration noise. After taking a certain action at the $t$-th slot, the system observes the instantaneous reward and the state in the $(t+1)$-th slot. The transition between the two slots, i.e., $\mathcal{T}(t)\triangleq\langle {\bf{s}}(t), {\bf{a}}(t),r(t), \allowbreak {\bf{s}}(t+1)\rangle$, is stored in a replay memory with the size of $|\mathcal{I}|$, where $\mathcal{I}$ is the set of time indices of transitions saved in the replay memory. After that, a batch of transitions are selected from the memory and used as the training samples to optimize the parameters of the NNs. The $i$-th transition in the selected batch is
$\mathcal{T}(t_i)=\langle{\bf{s}}(t_i), {\bf{a}}(t_i), r(t_i), {\bf{s}}(t_i+1)\rangle$, $t_i \in \mathcal{N}_{\rm tr}$, where $\mathcal{N}_{\rm tr} = \{t_1,\dots,t_{N_{\rm tr}}\}$ is the set of indices of transitions in the batch and $N_{\rm tr}$ is the batch size.

To optimize the parameters of the critic, we use Bellman equation \cite{sutton2011reinforcement},
\begin{equation}\label{eq:bellman}
\begin{aligned}
&Q\big({\bf{s}}(t_i),{\bf{a}}(t_i)\big) = \\
&\mathop{\mathbb{E}} \Big[ r(t_i) + \gamma Q\Big({\bf{s}}(t_i+1),\mu\big({\bf{s}}(t_i+1)|\theta^\mu\big)\Big) \Big |  {\bf{s}}(t_i), {\bf{a}}(t_i) \Big] .
\end{aligned}
\end{equation}
The realization of the right-hand side of \eqref{eq:bellman} in the $t_i$-th slot is defined as $y(t_i) \triangleq r(t_i) + \gamma Q\big({\bf{s}}(t_i+1),\mu\big({\bf{s}}(t_i+1)|\theta^\mu\big) | \theta^Q \big)$. To obtain an accurate approximation of the long-term reward, the DDPG algorithm minimizes the difference between $y(t_i)$ and $Q\big({\bf{s}}(t_i), {\bf{a}}(t_i) | \theta^Q \big)$. Thus, the parameters of the critic are optimized by minimizing the following loss function,
\begin{equation} \label{eq:loss_q}
\begin{aligned}
L(\theta^{Q}) = \frac{1}{N_{\rm tr}}\sum_{i=1}^{N_{\rm tr}}\Big[y(t_i)- Q\big({\bf{s}}(t_i), {\bf{a}}(t_i) | \theta^Q \big) \Big]^2  \ .
\end{aligned}
\end{equation}
Since the optimal policy maximizes the state-action value function, the loss function of the actor is defined as
\begin{equation} \label{eq:loss_a}
\begin{aligned}
L(\theta^{\mu}) = -\frac{1}{N_{\rm tr}}\sum_{i=1}^{N_{\rm tr}} Q\big({\bf{s}}(t_i),\mu\big({\bf{s}}(t_i) | \theta^{\mu}\big) | \theta^{Q} \big),
\end{aligned}
\end{equation}
which is minimized during training.

\subsubsection{DDPG with discrete actions}\label{subsec:discrete_action}
The original DDPG requires the actions to be continuous variables, which can be directly obtained from the continuous output of the actor. If the action space, $\mathcal{A}$, is discrete, we need to map the continuous output of the actor to a discrete action. With the method in \cite{qi2019deep}, we can find the closest valid action by
$\arg \min_{{\bf{a}}(t)\in \mathcal{A}} \left\| {\bf{a}}(t) -\mu({\bf{s}}(t)|\theta^\mu) \right\|_2$, 
where $\left\| {\bf{x}} \right\|_2$ is the $\ell_2$ norm of a vector ${\bf{x}}$.

\subsection{Problem Formulation}
In the straightforward implementation, we define the state, action, and reward by using the control signaling or observations that are directly available from 5G NR systems.

\subsubsection{Action of the scheduler}\label{subsec:scheduler_output}
The scheduler determines the numbers of RBs that will be allocated to different users. Thus, the action of the scheduler in the $t$-th slot is given by
\begin{equation}\label{eq:output_vector}
\begin{aligned}
{\bf{a}}(t) \  {\triangleq} \ \left[n_1(t),\dots,n_K(t)\right]^{\rm T},
\end{aligned}
\end{equation}
where ${n}_k(t)$ is the nearest integer to the $k$th element of $\mu\big({\bf{s}}(t)|\theta^\mu\big)$. If ${n}_k(t) = 0$, the $k$-th user will not be scheduled in the $t$-th slot.

\subsubsection{State of the network}\label{subsec:scheduler_input}
Since HoL delays and downlink SNRs are available at 5G BSs, we define the normalized state in the $t$-th slot as
\begin{equation}\label{eq:input_vector}
\begin{aligned}
{\bf{s}}(t) \  {\triangleq}
\left[ \frac{d_1(t)}{D_\text{max}},\dots,\frac{d_K(t)}{D_\text{max}},
\frac{\log\phi_1(t)}{\log\phi_\text{max}},\dots,\frac{\log\phi_K(t)}{\log\phi_\text{max}}
\right]^{\rm T} ,
\end{aligned}
\end{equation}
where $\phi_{\max}$ is the maximum SNR represented by the maximum channel quality indicator \cite{3gpp.38.214}.

\subsubsection{Reward of the scheduler}\label{subsec:scheduler_objective}
According to the QoS requirements in Section \ref{subsec:qos_tsn},
the total instantaneous reward in the $t$-th slot of the system is defined as the total number of packets successfully received by users in this slot,
\begin{equation}\label{eq:reward_total_model_free}
\begin{aligned}
{r}(t) \triangleq \sum_{k=1}^{K} {r}_k(t)\ ,
\end{aligned}
\end{equation}
where $r_k(t)$ is the number of packets received by the $k$-th user in the $t$-th slot, i.e.,
\begin{equation}\label{eq:reward_ue_model_free}
\begin{aligned}
{r}_k(t) =
    \begin{cases}
         \mathbf{1}_{D_\text{min} \leq d_k(t)\leq D_\text{max}} \cdot \mathbf{1}^{\text{dec}}_{k}(t) ,\ \text{if}\ x_k(t) =1\ , \\
        0\ ,\ \text{if}\ x_k(t) =0  \ ,
    \end{cases}
\end{aligned}
\end{equation}
where $\mathbf{1}_{D_\text{min} \leq d_k(t) \leq D_\text{max}}$ and $\mathbf{1}^{\text{dec}}_{k}(t)$ are two indicators. Specifically, if the packet is scheduled when $d_k(t) \in [D_\text{min}, D_\text{max}]$, then $ \allowbreak \mathbf{1}_{D_\text{min} \leq d_k(t)\leq D_\text{max}} \allowbreak = 1$, otherwise, $\mathbf{1}_{D_\text{min} \leq d_k(t)\leq D_\text{max}} = 0$. In the case that the packet is successfully decoded, $\mathbf{1}^{\text{dec}}_{k}(t) = 1$, otherwise, $\mathbf{1}^{\text{dec}}_{k}(t) = 0$.

Finally, the optimal control problem that maximizes the long-term reward of a scheduler can be formulated as follows,
\begin{equation}\label{eq:objective_long_term}
\begin{aligned}
\max_{\mu(\cdot|\theta^\mu)} \mathop{\mathbb{E}} \Big[ \sum_{i=0}^{\infty}\gamma^i r(t+i) \Big] \ .
\end{aligned}
\end{equation}

In this work, we only consider time-sensitive traffic.
To extend the work to 5G networks with other network services, we can formulate different scheduler design problems for other network services by changing the reward function. For example, we can use the long-term average rate as the reward function to maximize the fairness and the throughput of the system. However, in this case, PF is the optimal scheduler. Note that there is no need to use DRL or DDPG in the scenarios where the optimal schedulers are available. For time-sensitive traffic, the optimal scheduler is not available. Therefore, we use the learning-based methods to design the scheduler.

Furthermore, we can extend this work to 5G networks with multiple network services and one possible approach is to define the reward of the system as the weighted sum of utility functions of different services.

\section{Issues in the Straightforward Implementation of DDPG}\label{sec:challenges}
It seems that DDPG can be directly applied in scheduler design.
However, as to be shown in this section, there are some issues to be addressed in the straightforward implementation.
\subsection{Issues in Problem Formulation}\label{subsec:issue_problem_formulation}
\subsubsection{Large action space}
Denote the total number of RBs assigned to time-sensitive traffic by $N$. Thus, the possible number of RBs allocated to a user varies from $0$ to $N$. With $K$ users, the size of the action space in the straightforward implementation, $|\mathcal{A}|$, is $(N+1)^K$ based on \eqref{eq:output_vector}. 
In a 5G NR system, the value of $N$ can be higher than $100$ depending on the total bandwidth and the bandwidth of each RB. Thus, the action space could be extremely large. A reinforcement learning algorithm converges as the number of visits to each state-action pair approaches infinite \cite{sutton2011reinforcement}. When the action space is large, finding the optimal scheduler is nearly impossible in practice.

\subsubsection{Low flexibility of states}
Furthermore, the actor in the straightforward approach is a mapping from HoL delays and SNRs in \eqref{eq:input_vector} to the numbers of RBs in \eqref{eq:output_vector}. The total number of RBs assigned to time-sensitive traffic, $N$, the bandwidth of each RB, $W$, and the duration of each TTI, $\Delta^t$, are hidden variables that are not included in the input of DDPG, and hence are assumed to be constant. According to the standard of 3GPP, these hidden variables are flexible in 5G NR \cite{3gpp.38.214}. When $W$, $\Delta^t$ and $N$ become different, we need to train a new scheduler by using DDPG, which is inflexible and inefficient.

\subsubsection{Poor reliability evaluation}
In addition, for the time-sensitive traffic, the required reliability can be up to $99.999$\%. The long-term reward in \eqref{eq:objective_long_term} is linear with the reliability. Thus, by increasing the reliability of all users from $99.99$\% to $99.999$\%, the long-term reward only increases $0.01\%$. As a result, the gradient of \eqref{eq:loss_a} will be very small. Since the gradient descent method is used to update the parameters of the actor, the convergence time of the actor will be very long.

\subsection{Issues in Training Algorithm}\label{subsec:issue_training}
\subsubsection{Unaware of individual QoS}
The output of the critic is a single scalar value estimating the long-term reward of all users. Thus, it is not aware of the QoS of each user, and some users may suffer from poor QoS.

\subsubsection{Delayed reward/Sparse reward}
Due to the requirement on jitter, the scheduler receives positive rewards if packets are scheduled with HoL delays in $[D_{\min}, D_{\max}]$. When $d_k(t) < D_{\min}$, the instantaneous reward is $0$ no matter the packet is scheduled or not. In other words, the scheduler needs to take a series of actions to get a delayed non-zero reward. For example, only by taking the following actions, $n_k(t) = 0$ when $d_k(t) < D_{\min}$, and $n_k(t) > 0$ when $d_k(t)\in [D_{\min}, D_{\max}]$, the scheduler can receive a positive reward. Due to the fact that the non-zero rewards are sparse, such an issue is also referred to as sparse reward in \cite{sutton2011reinforcement}. Since the scheduler is not told which actions to take in order to get the non-zero reward, it is difficult for DDPG to learn the correct actions.

\subsubsection{Inaccurate critic at rarely visited state-action pairs}
To train the parameters of the critic, DDPG selects a batch of transitions from the replay memory with equal probabilities. For the state-action pairs that are visited with high frequency, they are more likely to be selected than the state-action pairs that are rarely visited. When the packet loss probability is small, the state-action pairs with packet losses are rarely visited and the critic is inaccurate at these state-action pairs. As a result, it is difficult for DDPG to achieve high reliability. 

\subsection{Issues in Online Implementation}\label{subsec:issue_online_implementation}
\begin{figure}[t]
\centering
\includegraphics[scale=0.7]{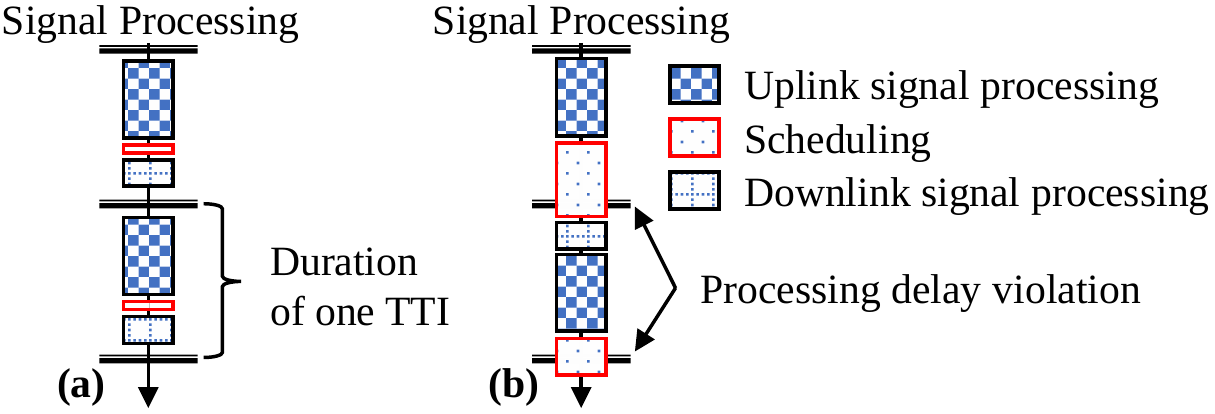}
\caption{Processing in the BS: (a) low-complexity scheduler, (b) high-complexity scheduler.}\label{fig:procssing}
\vspace{-0.2cm}
\end{figure}

\subsubsection{Poor initial QoS performance}
Since the parameters of the actor and critic are randomly initialized in DDPG, the QoS performance is poor during the first a few slots, which leads to high packet loss probability. Since the algorithm interacts with real-world networks, random initialization will cause severe QoS violations.

\subsubsection{Processing delay violation in each TTI}
The BS needs to perform scheduling and the baseband signal processing (i.e., decoding uplink packets and encoding downlink packets) within each TTI.
As shown in Fig.~\ref{fig:procssing}, if the processing of the feed-forward inference and the baseband signal processing is not finished in one TTI, radio signals cannot be transmitted in the assigned time slots, leading to radio link failures. We refer to this issue as processing delay violation.

\section{Theoretical DRL Framework}\label{sec:framework}
To address the issues in problem formulation, we propose a T-DRL framework to simplify the optimal control problem, where we exploit theoretical models and results to 1) reduce the action space, 2) generalize the state, and 3) evaluate the reliability.

\subsection{Theoretical Models and Results}\label{subsec:models}
The number of packets arrived at the queue of the $k$-th user in the $t$-th slot is denoted by $b_k(t)$.
For typical time-sensitive traffic, such as mission-critical IoTs and vehicle networks, the packet arrival processes follow Bernoulli processes, i.e., $b_k(t)\in\{0,1\}$ \cite{3GPP2012MTC,Hassan2013A,3GPP2017Scenarios} and the packet size is small, e.g., 20 or 32 bytes \cite{3GPP2017Scenarios}. We assume that with probability $p_k$, one packet arrives in the slot, $b_k(t)=1$. With probability $1-p_k$, no packet arrives in the slot, $b_k(t)=0$.

When transmitting a small packet, the required bandwidth is assumed to be smaller than the coherence bandwidth. 
We assume that the duration of each TTI is smaller than the channel coherence time.
Thus, the wireless channel is flat fading and quasi-static and the blocklength of channel codes is short. To transmit $L_k$ bits of data to the $k$-th user, the decoding error probability in the short blocklength regime can be accurately approximated by \cite{Yury2014Quasi}
\begin{equation}\label{eq:tx_error}
\begin{aligned}
\epsilon_k(t) \approx f_Q\Bigg(\frac{  -L_k\ln{2}+ {\Delta^t W n_k(t)}\ln\big[1+\phi_k(t)\big]}{\sqrt{\Delta^t W n_k(t) V_k(t)}}\Bigg) \ ,
\end{aligned}
\end{equation}
where $f_Q$ is Q-function, $W$ is the bandwidth of each RB,
$V_k(t)$ in (\ref{eq:tx_error}) is the channel dispersion defined as $V_k(t) = 1- {1}/{\big[1+\phi_k(t)\big]^2}$ \cite{Yury2014Quasi}.

To avoid long transmission delays and large jitters, retransmission cannot be used to improve reliability. In this case, if a packet is not successfully decoded by the user, it is lost. To achieve high reliability, the target decoding error probability of the $k$-th user should not exceed a threshold, i.e.,
\begin{align}
\epsilon_k(t) \leq \epsilon_{\max} \ .\label{eq:relia}
\end{align}
Since the decoding error probability in \eqref{eq:tx_error} decreases with the number of RBs, $n_k(t)$, the minimum number of RBs required to satisfy the constraint in \eqref{eq:relia}, denoted by $n^*_k(t)$, can be obtained via binary search \cite{She2018Joint}.

\subsection{Action Space Reduction}
If the $k$-th user is scheduled, the number of RBs required to guarantee the reliability in \eqref{eq:relia}, $n_k^*(t)$, can be obtained by substituting \eqref{eq:tx_error} into \eqref{eq:relia}. Thus, the scheduler only needs to determine which users to be scheduled. We define the action of the scheduler as
\begin{align}\label{eq:action_vector}
\hat{\bf{a}}(t) = [x_1(t),...,x_K(t)]^{\rm T} \ ,
\end{align}
where $x_k(t), k=1,\dots,K$, are binary variables that are obtained from the output of the actor according to Section \ref{subsec:discrete_action}. Thus, the number of possible actions is $2^K$, which is much smaller than that of the straightforward implementation, i.e., $(N+1)^K$, according to the definition of actions in \eqref{eq:output_vector}.
Given $\hat{\bf{a}}(t)$ in \eqref{eq:action_vector}, the number of RBs allocated to the $k$-th user can be obtained from the following expression,
\begin{equation}\label{eq:x_to_n_rb}
\begin{aligned}
n_k(t) =
    \begin{cases}
        x_k(t) n^*_k(t)\ , &\text{if}\ \sum_{k=1}^{K} x_k(t) n^*_k(t) \leq N \ , \\
        \Big\lfloor \frac{x_k(t) n^*_k(t)}{\sum_{k=1}^{K} x_k(t) n^*_k(t)} N \Big\rfloor , &\text{if}\ \sum_{k=1}^{K} x_k(t) n^*_k(t) > N \ . \\
    \end{cases}
\end{aligned}
\end{equation}

\subsection{Generalization of State}
To enable DRL in 5G NR with flexible configurations, we replace $\frac{\log\phi_k(t)}{\log\phi_\text{max}(t)}$ in \eqref{eq:input_vector} with $\frac{n^*_k(t)}{N}$, which is obtained from the theoretical formulas in \eqref{eq:tx_error} and \eqref{eq:relia} and depends on $\phi_k$, $N$, $W$ and $\Delta^t$. Based on the theoretical models and results, the generalized state of the system in the $t$-th slot is given by
\begin{equation}\label{eq:state_vector}
\begin{aligned}
\hat{\bf{s}}(t) {\triangleq} \left[ \frac{d_1(t)}{D_\text{max}},\dots,\frac{d_K(t)}{D_\text{max}},\frac{n^*_1(t)}{N},\dots,\frac{n^*_K(t)}{N} \right]^{\rm T} .
\end{aligned}
\end{equation}

\subsection{Reliability Evaluation}
From the definition of the decoding error probability in \eqref{eq:tx_error}, we have ${\mathbb{E}}[\mathbf{1}^{\text{dec}}_{k}(t)] = 1 - \epsilon_k(t)$. By replacing $\mathbf{1}^{\text{dec}}_{k}(t)$ in \eqref{eq:reward_ue_model_free} with $ 1 - \epsilon_k(t)$, the reward of the $k$-th user can be expressed as follows,
\begin{align}\label{eq:reward_ue_model_based}
\tilde{r}_k(t) =
    \begin{cases}
        \mathbf{1}_{D_\text{min} \leq d_k(t)\leq D_\text{max}}(1- \epsilon_k(t)),\ \text{if}\ x_k(t) =1,\\
        0,\ \text{if}\ x_k(t) =0  \ .
    \end{cases}
\end{align}
As mentioned in Section \ref{subsec:issue_problem_formulation}, when $\tilde{r}_k(t)$ is close to $1$, the training efficiency of DDPG is low. To handle this issue, we define the reward of the $k$-th user in the $t$-th slot as follows,
\begin{equation}\label{eq:reward_ue_log}
\begin{aligned}
\hat{r}_k(t) = -\log\big[1-\tilde{r}_k(t)\big]  \ .
\end{aligned}
\end{equation}
It is worth noting that the reward function in \eqref{eq:reward_ue_log} is not well defined in the straightforward implementation. This is because the reward in \eqref{eq:reward_ue_model_free}, $r_k(t)$, a binary number that may be equal to~$1$. With the definition in \eqref{eq:reward_ue_log}, the expectation of $\hat{r}_k(t)$ is more sensitive to the scheduling policy than the expectation of $\tilde{r}_k(t)$. For example, by increasing the reliability from $99$\% to $99.99$\%, ${\mathbb{E}}[\tilde{r}_k(t)]$ increases by $1$\%, but ${\mathbb{E}}[\hat{r}_k(t)]$ is doubled. 

The total reward is defined as  the summation of the rewards of all users, i.e.,
\begin{equation}\label{eq:reward_total_model_based}
\begin{aligned}
\hat{r}(t) = \sum_{k=1}^{K} \hat{r}_k(t)\ .
\end{aligned}
\end{equation}

\subsection{Markov Property}
To apply DRL, we need to prove that transitions of the system follow an MDP. Otherwise, DRL algorithms may not converge \cite{sutton2011reinforcement}. By assuming that the wireless channel fading is Markovian~\cite{WirelessCom}, the Markov property holds for the scheduler design problem (See proof in Appendix).

\section{Knowledge-assisted DDPG}\label{sec:k-ddpg}
To address the issues in the training phase of the straightforward implementation, we propose K-DDPG by exploiting expert knowledge, which is formally defined as the design principles and insights from human experts. Specifically, for the scheduler design problem, the expert knowledge includes 1) the rewards of multiple users, 2) the target scheduling policy and 3) the importance of transitions. With the help of knowledge, K-DDPG can improve the QoS of each user and reduce the convergence time.

\subsection{Multi-head Critic for Individual QoS Evaluation}
\begin{figure}[t]
\centering
\includegraphics[scale=0.8]{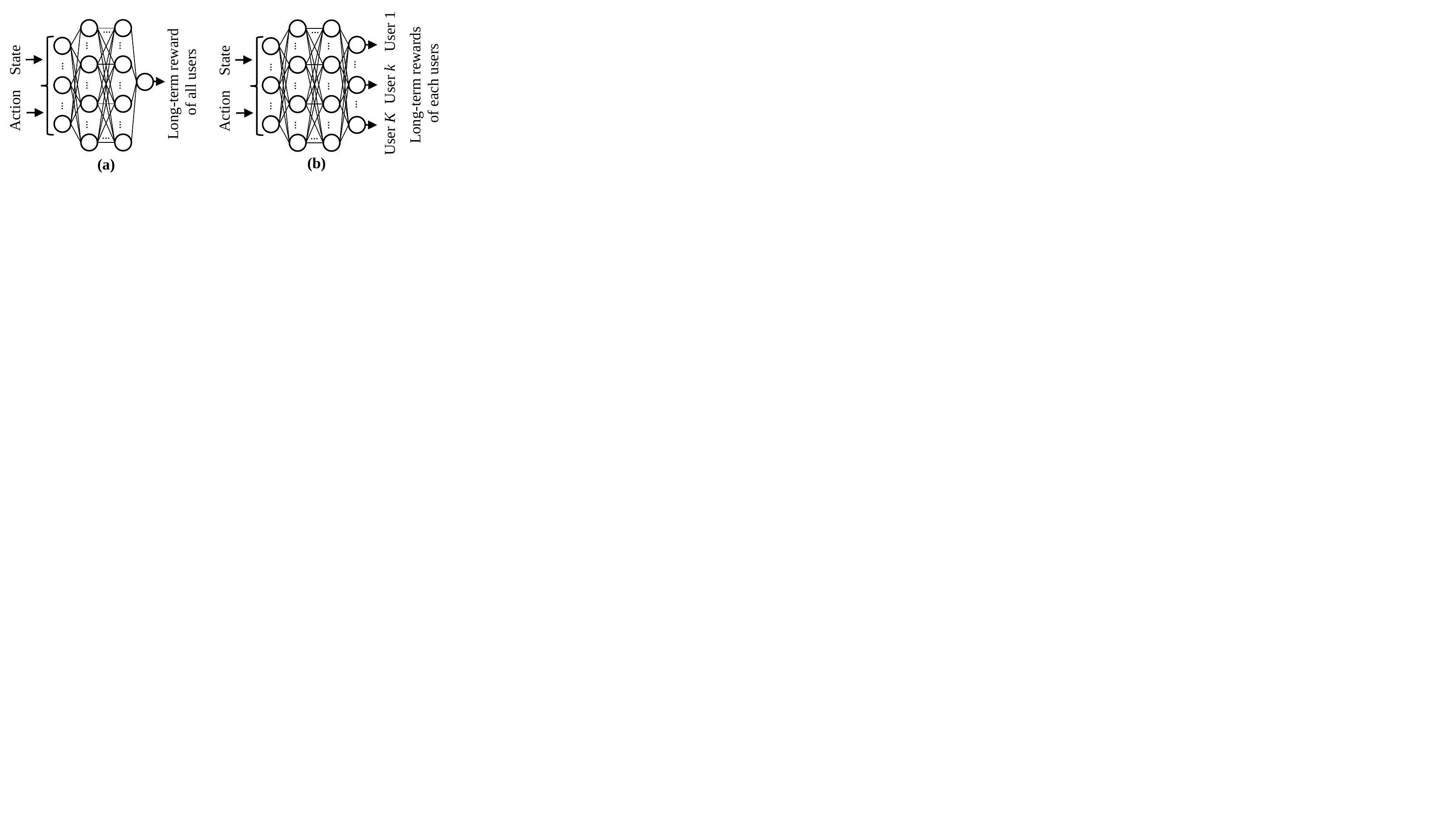}
\caption{Illustration of (a) single-head critic and (b) multi-head critic.}
\label{fig:critics}
\vspace{-0.5cm}
\end{figure}
The single-head critic in Fig.~\ref{fig:critics} is not aware of the reward of each component in the system, e.g., the QoS of each user, $\hat{r}_k(t)$, $k=1,\dots,K$. Since there are multiple users, the total long-term reward is the summation of the long-term reward of each user, according to \eqref{eq:reward_total_model_based}. Based on the knowledge of the reward structure, we decompose the reward into $K$ components for $K$ users.
We denote the rewards of all users at the $t$-th slot as
\begin{equation}\label{eq:reward_vector}
\begin{aligned}
\hat{\bf{r}}_k(t) \triangleq [\hat{r}_1(t),\dots,\hat{r}_K(t)] \ .
\end{aligned}
\end{equation}
The decomposed long-term rewards are approximated by the state-action value function denoted by $Q_1({\bf{s}},{\bf{a}}),\dots, Q_K({\bf{s}},{\bf{a}})$, where $Q_k({\bf{s}},{\bf{a}})$ is the state-action value function of the $k$-th user, defined as follows,
\begin{equation}\label{eq:q_pi_k}
\begin{aligned}
&Q_k(\hat{\bf{s}}(t),\hat{\bf{a}}(t))=\\
&\mathop{\mathbb{E}} \Big[ \sum_{i=0}^{\infty}\gamma^i \hat{r}_k(t+i) \Big| \hat{\bf{s}}(t), \hat{\bf{a}}(t), \hat{\bf{a}}(t') = \mu\big(\hat{\bf{s}}(t')|\theta^{\mu}\big), \forall t'>t\Big] .
\end{aligned}
\end{equation}

\subsection{Reward Shaping for Instantaneous Feedback}

\begin{figure}[t]
\centering
\includegraphics[scale=0.75]{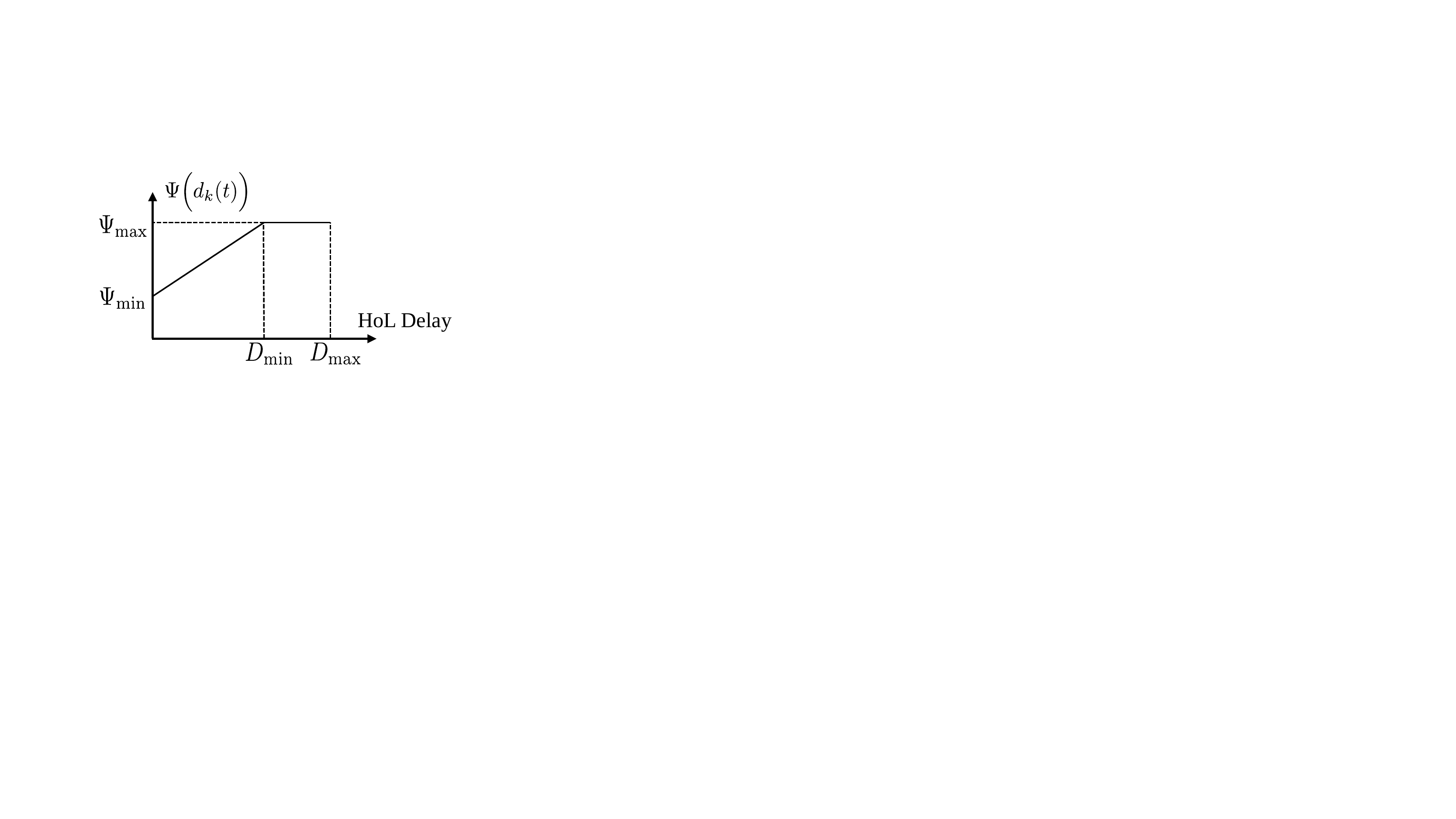}
\caption{Illustration of potential function for reward shaping.}
\label{fig:potential_function}
\vspace{-0.2cm}
\end{figure}

Exploring the optimal policy with delayed rewards/sparse rewards is time consuming. To handle this issue, we apply \textit{reward shaping} \cite{ng1999policy} to generate non-zero instantaneous feedback in each slot. According to the requirement on jitter, a target scheduling policy should only schedule users with $d_k(t) \in [D_{\min}, D_{\max}]$. For users with $d_k(t) < D_{\min}$, they should not be scheduled. Based on the knowledge of the target scheduling policy, we define a potential function, $\Psi(d_k(t))$, which generates non-zero instantaneous reward according to
\begin{align}\label{eq:shaping}
\dot{r}_k(t) = \hat{r}_k(t) - \Psi\big(d_k(t)\big) + \gamma\Psi\big(d_k(t+1)\big) \ ,
\end{align}
which is the shaped reward of the $k$-th user.
To illustrate the relation between the instantaneous feedback and the potential function, we considered an example in Fig.~\ref{fig:potential_function}, where the potential function increases linearly with $d_k(t)$ when $d_k(t) < D_{\min}$. When $d_k(t) < D_{\min}$, $\hat{r}_k(t) = 0$ no matter whether the user is scheduled or not. If the user is not scheduled, $d_k(t+1) = d_k(t)+1$, then $- \Psi\big(d_k(t)\big) + \gamma\Psi\big(d_k(t+1)\big) > 0$ since $\gamma$ is closed to $1$. In other words, the scheduler will receive a positive instantaneous reward. On the other hand, if the user is scheduled, then $d_k(t+1) < d_k(t)$ and $- \Psi\big(d_k(t)\big) + \gamma\Psi\big(d_k(t+1)\big) < 0$. In this case, the scheduler will receive a negative instantaneous reward.

When reward shaping is used in reinforcement learning, the state-action value function, denoted by $\dot{Q}_k(\hat{\bf{s}}(t),\hat{\bf{a}}(t))$, will be different from the original $Q_k(\hat{\bf{s}}(t),\hat{\bf{a}}(t))$. 
The Bellman equation can be re-expressed as follows,
\begin{equation}\label{eq:multi_head_critic_bellman}
\begin{aligned}
&\dot{Q}_k\left(\hat{\bf{s}}(t),\hat{\bf{a}}(t)\right) = \\
&\mathop{\mathbb{E}} \big[  \dot{r}(t) + \gamma \dot{Q}_k\Big(\hat{\bf{s}}(t+1),\mu (\hat{\bf{s}}(t+1)|\theta^\mu)\Big) \Big|  \hat{\bf{s}}(t), \hat{\bf{a}}(t)\big] \ ,
\end{aligned}
\end{equation}
where $k = 1\dots,K$. By substituting \eqref{eq:shaping} into \eqref{eq:multi_head_critic_bellman}, we can derive that (See the details in \cite{ng1999policy}.)
\begin{align}\label{eq:q_pi_k_rs}
\dot{Q}_k(\hat{\bf{s}}(t),\hat{\bf{a}}(t))=Q_k(\hat{\bf{s}}(t),\hat{\bf{a}}(t)) - \Psi\big(d_k(t)\big) \ .
\end{align}
Since $\Psi\big(d_k(t)\big)$ does not  depend on the action to be taken, the actor that maximizes $\dot{Q}_k(\hat{\bf{s}}(t),\hat{\bf{a}}(t))$ is the same as the actor that maximizes $Q_k(\hat{\bf{s}}(t),\hat{\bf{a}}(t))$, i.e.,
\begin{equation}\label{eq:rs}
\begin{aligned}
&\arg\max\limits_{{{\mu(\cdot|\theta ^\mu)} }} \dot{Q}_k(\hat{\bf{s}}(t),\mu(\hat{\bf{s}}(t)|\theta ^\mu))\\
=&\arg\max\limits_{{{\mu(\cdot|\theta ^\mu)} }} Q_k(\hat{\bf{s}}(t),\mu(\hat{\bf{s}}(t)|\theta ^\mu)) - \Psi\big(d_k(t)\big)\\
=&\arg\max\limits_{{{\mu(\cdot|\theta ^\mu)} }} Q_k(\hat{\bf{s}}(t),\mu(\hat{\bf{s}}(t)|\theta ^\mu)) \ .
\end{aligned}
\end{equation}
Therefore, the optimal actor does not change with the potential function \cite{ng1999policy}.

After reward shaping, the multi-dimensional state-action value function of the $K$ users is defined as
${\bf{Q}}\left(\hat{\bf{s}}(t),\hat{\bf{a}}(t)\right) \triangleq [\dot{Q}_1\left(\hat{\bf{s}}(t),\hat{\bf{a}}(t)\right),\dots,\allowbreak \dot{Q}_K\left(\hat{\bf{s}}(t),\hat{\bf{a}}(t)\right)]^{\rm T}$ \cite{van2017hybrid}.
As illustrated in Fig.~\ref{fig:critics}b, ${\bf{Q}}\left(\hat{\bf{s}}(t),\hat{\bf{a}}(t)\right)$ is approximated by a multi-head critic,
\begin{equation}\label{eq:multi_head_critic_approximation}
\begin{aligned}
&{\bf{Q}}\big(\hat{\bf{s}}(t),\hat{\bf{a}}(t)|\Theta^q\big)\triangleq \\
& \left[\dot{Q}_1\big(\hat{\bf{s}}(t),\hat{\bf{a}}(t)|\Theta^q\big),\dots,\dot{Q}_K\big(\hat{\bf{s}}(t),\hat{\bf{a}}(t)|\Theta^q\big)\right]^{\rm T} \ ,
\end{aligned}
\end{equation}
which is a NN with parameters $\Theta^q$.

\subsection{Importance Sampling}
To train the critic, the original DDPG selects a batch of training samples from $|\mathcal{I}|$ transitions in the replay memory. All transitions will be selected with the same probability, $1/|\mathcal{I}|$. However, transitions are not with the same importance. In scheduler design, transitions with higher approximation errors of the state-action value function or with more packet losses are more important than the other transitions. Specifically, we define a weight, $w(t)$, of transition $\hat{\mathcal{T}}(t), t \in \mathcal{I}$. The probability that transition  $\hat{\mathcal{T}}(t)$ will be selected is given by \cite{schaul2015prioritized}
\begin{equation}\label{eq:importance_sampling}
\begin{aligned}
p_{\rm tr}(t) = \frac{{w(t)}}{\sum_{i\in\mathcal{I}}{w(i)}} \ .
\end{aligned}
\end{equation}
We set the initial weight of the transition generated in the $t$-th time slot, $\hat{\mathcal{T}}(t)$, as maximum weight of all transitions that have been stored in the replay memory during the previous $(t-1)$ slots, $w(t) = \max_{i\in\mathcal{I}}{w(i)}$. For the transition generated in the first slot, the weight is set as a small positive number in order to avoid zero weight.

Based on the selected batch, we first update the weights of transitions in the batch based on the approximation error of the critic and the number of packet losses, i.e., 
\begin{equation} \label{eq:weight_update}
\begin{aligned}
&w(t_i) \leftarrow  \sum_{k=1}^{K}\Big\{\big[ \dot{y}_k(t_i) -\dot{Q}_k\big(\hat{\bf{s}}(t_i), \hat{\bf{a}}(t_i) | \Theta^q \big)  \big]^2\\
&\cdot  \big[ 1+(1-x_k)\mathbf{1}_{d_k(t) = D_\text{max}}+x_k\mathbf{1}_{d_k(t) \notin [D_\text{min}, D_\text{max}]} \big] \Big\}
\ ,
\end{aligned}
\end{equation} 
where $\dot{y}_k(t_i)=\dot{r}_k(t_i) + \gamma \dot{Q}_k\Big(\hat{\bf{s}}(t_i+1), \mu\big(\hat{\bf{s}}(t_i+1) | \theta^\mu\big)|\Theta^q\Big)$ is the realization of the right-hand side of \eqref{eq:multi_head_critic_bellman} in the $t_i$-th slot.
The first part in \eqref{eq:weight_update} is the approximation error of the critic.
The second part in \eqref{eq:weight_update} depends on the number of packet losses. Specifically, if the $k$-th user is not scheduled when $d_k(t) = D_{\max}$ or is scheduled when $d_k(t)\notin[D_{\min},D_{\max}]$, there is a packet loss. If there is a packet loss, the second part equals to $2$. Otherwise, it equals to $1$.

To optimize the parameters of the multi-head critic, we minimize the following loss function,\footnote{Since transitions are not selected with the same probability from the replay memory, $\frac{1}{N_{\rm tr}}\sum_{i=1}^{N_{\rm tr}}  \sum_{k=1}^{K} \big[ \dot{y}(t_i) -\dot{Q}_k\big(\hat{\bf{s}}(t_i), \hat{\bf{a}}(t_i) | \Theta^q \big) \big]^2$ is no longer the average of the approximation error of the state-action value function.}
\begin{equation}\label{eq:loss_multi-head_q}
\begin{aligned}
L(\Theta^{q}) =
\frac{1}{N_{\rm tr}}\sum_{i=1}^{N_{\rm tr}}  u(t_i) \sum_{k=1}^{K} \big[ \dot{y}_k(t_i) -\dot{Q}_k\big(\hat{\bf{s}}(t_i), \hat{\bf{a}}(t_i) | \Theta^q \big) \big]^2 \ ,  
\end{aligned}
\end{equation} 
where the co-efficient $u(t_i)$ is defined as follows,
\begin{equation}\label{eq:bias_correction}
\begin{aligned}
u(t_i) = \frac{1}{p_{\rm tr}(t_i)|\mathcal{I}|} \ ,
\end{aligned}
\end{equation}
which corrects the bias caused by importance sampling.

To find the optimal actor that maximizes the total state-action value, $\sum_{k=1}^{K}\dot{Q}_k\big(\hat{\bf{s}}(t),\hat{\bf{a}}(t)\big)$ \cite{van2017hybrid}, we optimize the parameters of the actor to minimize the following loss function,
\begin{equation}\label{eq:loss_multi-head_a}
\begin{aligned}
L(\theta^{\mu}) &= \mathop{\mathbb{E}} \Big[ - \sum_{k=1}^{K} \dot{Q}_k\big(\hat{\bf{s}}(t),\hat{\bf{a}}(t)\big) \Big] \\ 
&=-\frac{1}{N_{\rm tr}}\sum_{i=1}^{N_{\rm tr}} u(t_i) \sum_{k=1}^{K} \dot{Q}_k\Big(\hat{\bf{s}}(t_i),\mu\big(\hat{\bf{s}}(t_i)|\theta^\mu\big)|\Theta^q  \Big) \ .
\end{aligned}
\end{equation}
The temporal copies of the $\Theta^{q}$ and $\theta^\mu$ are denoted by $\hat{\Theta}^{q}$ and $\hat{\theta}^\mu$, respectively. We first optimize $\hat{\Theta}^{q}$ and $\hat{\theta}^\mu$ and then update $\Theta^{q}$ and $\theta^\mu$ ``softly".
From \eqref{eq:loss_multi-head_q} and \eqref{eq:loss_multi-head_a}, we can derive the gradients of $L(\hat{\Theta}^{q})$ and $L(\hat{\theta}^\mu)$, respectively, i.e.,

\begin{align}
&\nabla_{\hat{\Theta}^{q}} L(\hat{\Theta}^{q}) 
=\frac{1}{N_{\rm tr}}\sum_{i=1}^{N_{\rm tr}} u(t_i)\sum_{k=1}^{K}\Bigg\{2 \Big[\dot{y}_k(t)- \nonumber\\ 
& \dot{Q}_k\big(\hat{\bf{s}}(t_i), \hat{\bf{a}}(t_i) | \hat{\Theta}^q \big) \Big]    \times \nabla_{\hat{\Theta}^{q}}  \dot{Q}_k\big(\hat{\bf{s}}(t_i), \hat{\bf{a}}(t_i) |\hat{\Theta}^q \big)\Bigg\} \label{eq:loss_multi-head_q_gradient},
\end{align}
\begin{align}
&\nabla_{\hat{\theta}^{\mu}}  L(\hat{\theta}^{\mu})  = -\frac{1}{N_{\rm tr}}\sum_{i=1}^{N_{\rm tr}}  u(t_i)\sum_{k=1}^{K}\Big[ \nonumber\\
&\nabla_{\bf{a}} \dot{Q}_k\big(\hat{\bf{s}}(t_i), {\bf{a}} | \hat{\Theta}^{q} \big) |_{ {\bf{a}}=\mu(\hat{\bf{s}}(t_i) |\hat{\theta}^{\mu})} \times \nabla_{\hat{\theta}^{\mu}}\mu(\hat{\bf{s}}(t_i) |\hat{\theta}^{\mu})\Big] \label{eq:loss_multi-head_a_gradient}.
\end{align}
The pseudo-code of K-DDPG is provided in Algorithm \ref{alg:ddpg_multi-head}.

\begin{algorithm}[!ht]
\caption{Knowledge-assisted DDPG.}\label{alg:ddpg_multi-head}
\begin{algorithmic}[1]
\STATE Initialize the parameters of the NNs, $\Theta^q$ and $\theta^\mu$.
\STATE Initialize temporal copies of the parameters: $\hat{\Theta}^q \leftarrow \Theta^q$ and $\hat{\theta}^\mu \leftarrow \theta^\mu$.
\STATE Initialize a replay memory with a size of $I$.
\FOR{episode $m$ = $1,\dots,M$}
    \STATE Set the system to an initial state, e.g., set queues of users as empty.
    \FOR{$t=(m-1)T+1,\dots,mT$}
        \STATE Observe state $\hat{\bf{s}}(t)$.
        \STATE Generate action from $\hat{\bf{a}}(t)=\mu(\hat{\bf{s}}(t)|\theta^{\mu})+\mathcal{N}(t)$, and execute the action.
        \STATE Evaluate reward $\hat{\bf{r}}(t)$ from \eqref{eq:tx_error}, \eqref{eq:reward_ue_model_based} and \eqref{eq:reward_ue_log}, and observe the next state $\hat{\bf{s}}(t+1)$.
        \STATE Save transition $\hat{\mathcal{T}}(t)=\langle \hat{\bf{s}}(t), \hat{\bf{a}}(t),\hat{\bf{r}}(t), \hat{\bf{s}}(t+1)\rangle$ and its weight $w(t)=\max_{i\in\mathcal{I}}{w(i)}$.
        \STATE Select $N_{\rm tr}$ transitions as a batch of training samples based on \eqref{eq:importance_sampling}.
        \STATE Update weights of selected transitions based on \eqref{eq:weight_update}.
        \STATE Compute $\nabla_{\hat{\Theta}^{q}} L(\hat{\Theta}^{q})$ and  $\nabla_{\hat{\theta}^{\mu}}  L(\hat{\theta}^{\mu})$ from  \eqref{eq:loss_multi-head_q_gradient} and \eqref{eq:loss_multi-head_a_gradient}, respectively.
        \STATE Optimize $\hat{\Theta}^Q$ and $\hat{\theta}^\mu$ with the SGD algorithm.
        \STATE Update $\Theta^{q}$ and $\theta^{\mu}$ based on $\hat{\Theta}^q$ and $\hat{\theta}^\mu$: \\ 
        \qquad $\Theta^{q} \leftarrow (1-\tau) \Theta^{q}  + \tau \hat{\Theta}^q \ ; \
        \theta^{\mu} \leftarrow (1-\tau) \theta^{\mu}  + \tau \hat{\theta}^\mu\ .$
    \ENDFOR
\ENDFOR
\STATE  Return $\Theta^q$ and $\theta^\mu$ for online fine-tuning.
\end{algorithmic}
\end{algorithm}

\section{Online DDPG Architecture}\label{sec:oneline_archiecture}
\begin{figure}[!hbt]
\centering
\includegraphics[width=0.99\columnwidth]{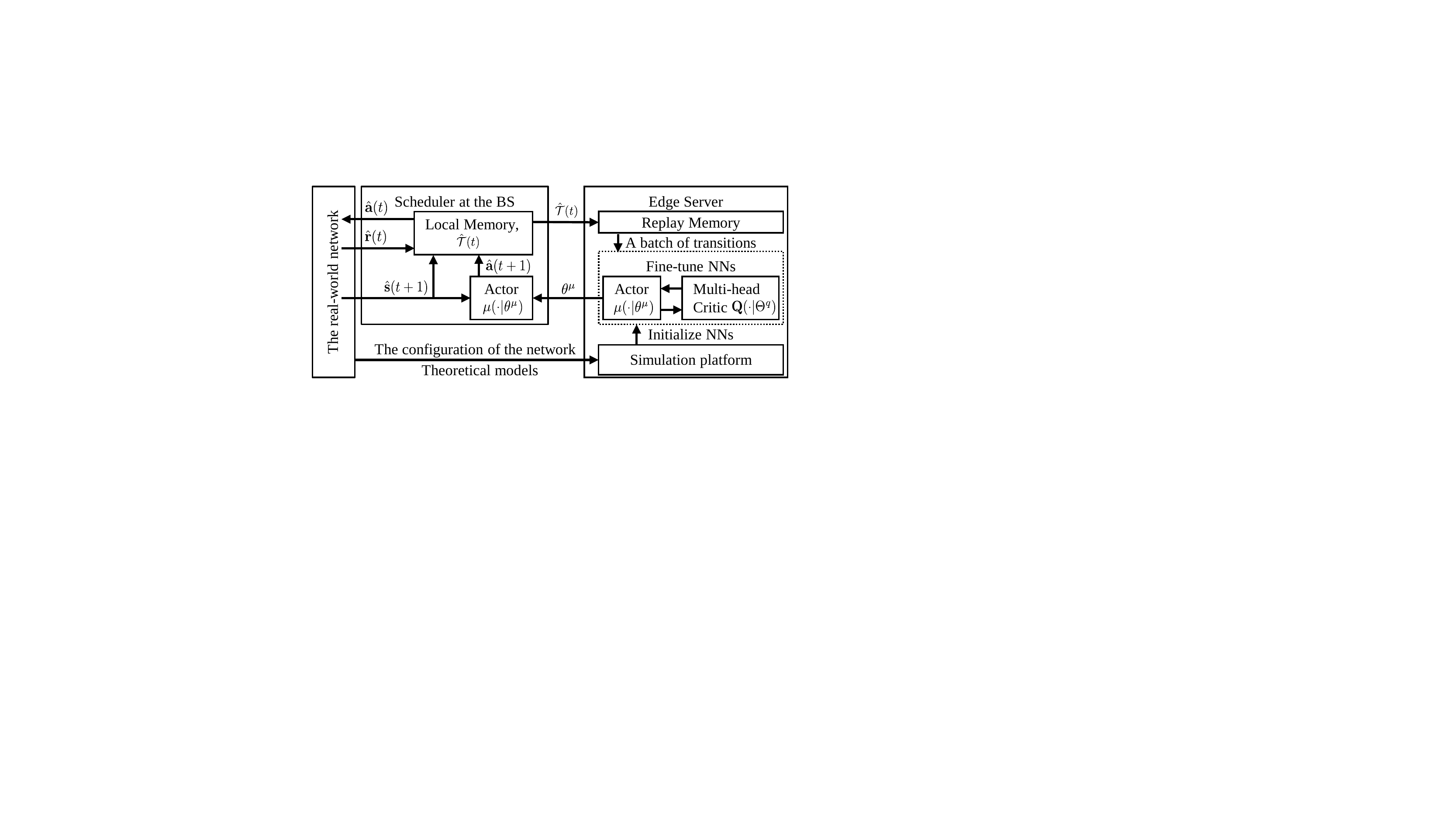}
\caption{Proposed online DDPG architecture.}
\label{fig:ddrl_achitecture_detail}
\vspace{-0.2cm}
\end{figure}

In this section, we address the issues in the real-world implementation of DDPG by proposing an architecture for online training and inference.
As shown in Fig.~\ref{fig:ddrl_achitecture_detail}, the online DDPG architecture includes the scheduler at the BS and an edge server.

\subsection{Off-line Initialization}\label{subsec:offline}
Before executing DDPG in the online architecture, we need to initialize the actor and the critic off-line in a simulation platform, which is built upon the configurations of the real-world network and the theoretical models. The basic idea is to generate transitions from the simulation platform and train the actor and the critic by using Algorithm \ref{alg:ddpg_multi-head}.
Considering that the simulation is not exactly the same as the real-world network, the actor and the critic are fine-tuned in the online architecture, which is introduced in the sequel.

\subsection{Scheduler at the BS}
\begin{figure}[!hbt]
\centering
\includegraphics[scale=0.85]{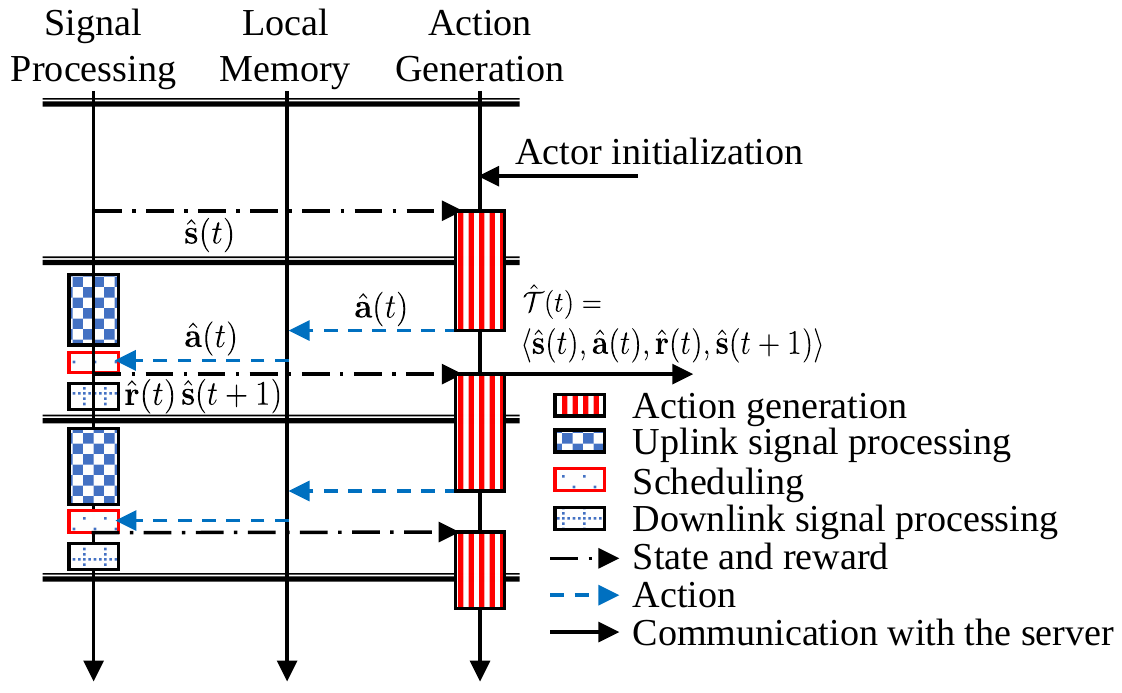}
\caption{Parallel processing in the BS.}
\label{fig:parallel_nn_fp}
\vspace{-0.2cm}
\end{figure}

After off-line initialization, the BS fetches the parameters of the actor, $\theta^\mu$. In the $t$-th TTI, the BS observes the state, $\hat{\bf{s}}(t)$, and generates an action according to the actor, $\hat{\bf{a}}(t) = \mu(\hat{\bf{s}}(t)|\theta^\mu)$.
To avoid processing delay violation mentioned in Section \ref{subsec:issue_online_implementation}, tasks for the action generation and the baseband signal processing are executed in parallel, as shown in Fig.~\ref{fig:parallel_nn_fp}.
The generated action is saved in a local memory before it is executed in scheduling. The numbers of RBs allocated to the scheduled users are given by \eqref{eq:x_to_n_rb}. After the action is executed, the BS computes the reward from \eqref{eq:reward_ue_log} and observes the state in the next TTI, $\hat{\bf{s}}(t+1)$. Finally, the transition, $\hat{\mathcal{T}}(t)=\langle\hat{\bf{s}}(t), \hat{\bf{a}}(t),\hat{\bf{r}}(t),\hat{\bf{s}}(t+1)\rangle$, is uploaded to the edge server and saved in the replay memory.

\subsection{Online Training in the Edge Server}\label{subsec:online_training}
In the edge server, the actor and the critic are initialized with the method in Section \ref{subsec:offline}.
Then, the server fine-tunes the actor and the critic by using transitions from the real-world scheduler at the BS.
Specifically, this is achieved by executing lines 11-15 of Algorithm \ref{alg:ddpg_multi-head}, iteratively.
Once the actor and the critic are updated in each iteration, the parameters of the actor are sent to the scheduler. In order to enable the exploration in the real-world network, the server can add a noise in the \textit{parameter space} of the actor according to $\theta^\mu  \leftarrow \theta^\mu \cdot (1+ \mathbb{N}(0,v^2) \cdot e ^ {-\lambda t \Delta^t})$, where $\mathbb{N}(0,v^2) \cdot e ^ {-\lambda t \Delta^t}$ are Gaussian noises that attenuates over time, $v$ is the variance of the noise and $\lambda$ is the attenuation rate \cite{plappert2018parameter}.

As shown in the online architecture in Fig. \ref{fig:ddrl_achitecture_detail}, the rewards of users, $\hat{\bf{r}}(t)$, are uploaded to the edge server in each slot by the BS, where three kinds of knowledge are exploited in the K-DDPG algorithm. First, based on the knowledge that the QoS of the whole system depends on the QoS of each user, we use the multi-head critic in the edge server to approximate the long-term rewards of different users. Second, the shaped rewards, $\dot{\bf{r}}(t)$, is obtained from \eqref{eq:shaping}, where the form of the potential function $\Psi(d_k(t))$ is designed by human experts based on their understanding of the knowledge of the target scheduling policy.
Third, the knowledge of the importance of transitions is updated according to  \eqref{eq:weight_update}, where the weight of each transition depends on the approximation errors of the value function and the number of packet losses.

\section{Simulation Results}\label{sec:sim_results}
\begin{table*}[t]
\vspace{0.2cm}\small
\caption{Simulation Setup}
\label{tab:system_param}
\begin{minipage}{\textwidth}
\begin{center}
\begin{tabular}{|p{6cm}|p{2cm}|p{4.5cm}|p{1.75cm}|}
\hline
\multicolumn{2}{|c|}{System setup \cite{3GPP2017Scenarios,3GPP2017Agree}} &  \multicolumn{2}{|c|}{Learning setup}  \\\hline
BS transmit power spectrum density $\mathbf{P}_{\text{max}}$& $20$~dBm/Hz  & Exploration parameters $\sigma$, $\delta$& $1$, $0.4$ \\\hline
Noise power spectrum density $\mathbf{N}_0$& $-90$~dBm/Hz   & Actor learning rate & $10^{-3}$\\\hline
Time slot duration (one TTI) $\Delta^t$ & $125$~us   &  Critic learning rate & $10^{-3}$\\\hline
Bandwidth of a RB $W$ & $180$~kHz  &  Soft-updating rate $\tau$ & $10^{-3}$\\\hline
Packet size $L_k$ & $32$~bytes &  Replay memory size $|\mathcal{I}|$ & $10000$\\\hline
Packet arrival probability $p_k$ & $10\%$  & Batch size $N_{\rm tr}$ & $20$\\\hline
Required decoding error probability $\epsilon_\text{max}$ & $10^{-5}$  & Time slots per episode $T$ & 200\\\hline
Timeliness requirement $[D_\text{min},D_\text{max}]$ &$[5,7]$ & Potential function $\Psi_{\min}$, $\Psi_{\max}$  & $0$, $1$  \\\hline
Maximum SNR $\log\phi_{\max}$ & $3.8$ & Discount factor $\gamma$ & $0.9$  \\\hline
\end{tabular}
\end{center}
\end{minipage}
\vspace{-0.2cm}
\end{table*}
\begin{figure*}[!ht]
\centering
\includegraphics[scale=0.8]{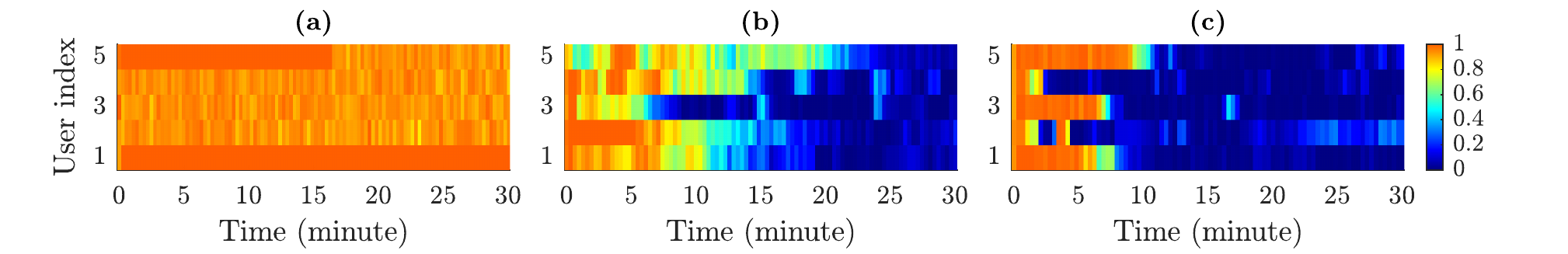}
\caption{Packet loss probabilities, where $K=5$ and $N=50$. (a) Straightforward implementation of DDPG, (b) DDPG in T-DRL framework, (c) K-DDPG in T-DRL framework.}
\label{fig:packet_loss_K5}
\vspace{-0.2cm}
\end{figure*}
\begin{figure*}[!ht]
\centering
\includegraphics[scale=0.8]{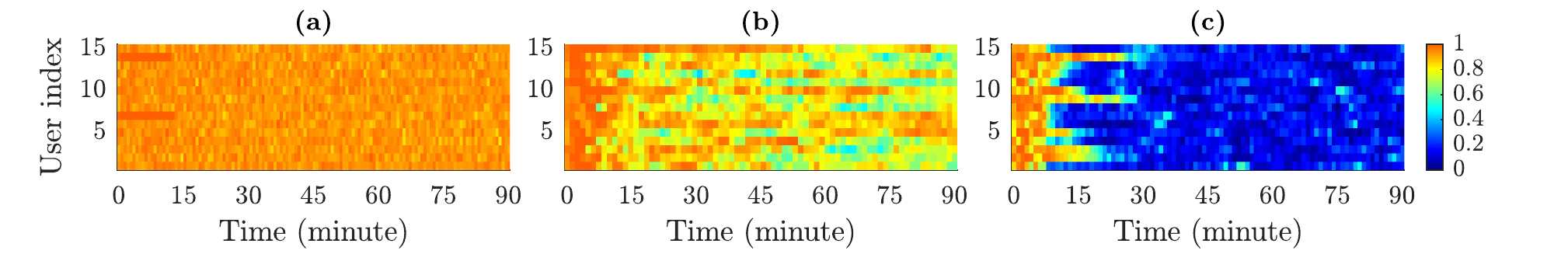}
\caption{Packet loss probabilities, where $K=15$ and $N=50$. (a) Straightforward implementation of DDPG, (b) DDPG in the T-DRL framework, (c) K-DDPG in T-DRL framework.}
\label{fig:packet_loss_K15}
\vspace{-0.2cm}
\end{figure*}

\subsection{Simulation Platform}
In the simulation platform, users randomly move with the velocity $5$~meters/second in a cell with a radius of $100$ meters. The channel models are the same as that in Section \ref{subsec:models}. The path loss model is $45+30\log(l)$ dB, where $l$ is the distance between a user and the BS in meter.
At the beginning of each episode, we set each user at a random position in the cell.
We assume that BS is in a factory and the small-scale channel gain follows a Rician distribution \cite{WirelessCom}. The ratio of the average power in the Line-of-Sight path to that in the Non-Line-of-Sight paths is set as $0.6$. We consider a discrete-time channel model in the simulation. Given the small-scale channel gain in the current slot, with probability $80$\%, it remains the same in the next slot; with probability $20$\%, it varies according to the Rician fading. 

For hyper-parameters in DRL (i.e., exploration rates, learning rates, and soft-updating rate in Table \ref{tab:system_param}), we tried different values and choose the best ones in this section.
Both the actor and the critic have one input layer, one output layer, and two hidden layers. The number of neurons in each layer depends on the number of users. Specifically, the dimensions of the four layers of the actor are $2K$, $20K$, $20K$ and $K$, respectively. The activation functions of the two hidden layers are RELU function. To ensure the output of the actor lies in  $[0,1]$, the activation function of the output layer is $\frac{1}{2}\text{TANH}(\cdot)+\frac{1}{2}$. For the critic, the dimensions of the four layers are $3K$, $30K$, $30K$, and $K$, respectively. RELU function is used as the activation functions of the two hidden layers, and no activation function is used in the output layer. In the simulation, the exploration noise, $\mathcal{N}(t) \triangleq [\mathcal{N}_1(t),\dots,\mathcal{N}_K(t)]^{\rm T}$, is added to the output of the actor, where $\mathcal{N}_k(t) = \mathcal{N}_k(t-1) +  \delta \cdot \mathbb{N}(0,\sigma^2)$. 
$\mathbb{N}(0,\sigma^2)$ is a Gaussian variable with zero mean and variance $\sigma^2$. The parameter $\delta$ is an adjustable exploration rate.
The simulation setup is summarized in Table \ref{tab:system_param}, unless mentioned otherwise. 

\subsection{Performance of the T-DRL Framework and the K-DDPG Algorithm}

To illustrate the benefits of the T-DRL framework and the K-DDPG algorithms, Figs. \ref{fig:packet_loss_K5} and \ref{fig:packet_loss_K15} show the packet loss probabilities during off-line training in the simulation. The packet loss probabilities are measured every $5$ episodes. When the number of users is small ($K=5$ in Fig.~\ref{fig:packet_loss_K5}), DDPG converges after $25$ minutes in the T-DRL framework, while the straightforward implementation of DDPG does not converge to a policy with low packet loss probabilities. With different types of expert knowledge of the scheduler design problem, K-DDPG can further reduce $50$\% of convergence time compared with DDPG (in the T-DRL framework). When the number of users is large ($K=15$ in Fig.~\ref{fig:packet_loss_K15}), DDPG can hardly obtain a satisfactory scheduler without the assistance of knowledge. The results in Figs. \ref{fig:packet_loss_K5} and \ref{fig:packet_loss_K15} indicate that by applying K-DDPG in T-DRL framework, the scheduler learns faster than the cases without the knowledge or theoretical models.

\begin{figure}[t]
\centering
\begin{subfigure}{.8\columnwidth}
\centering
\includegraphics[scale=0.7]{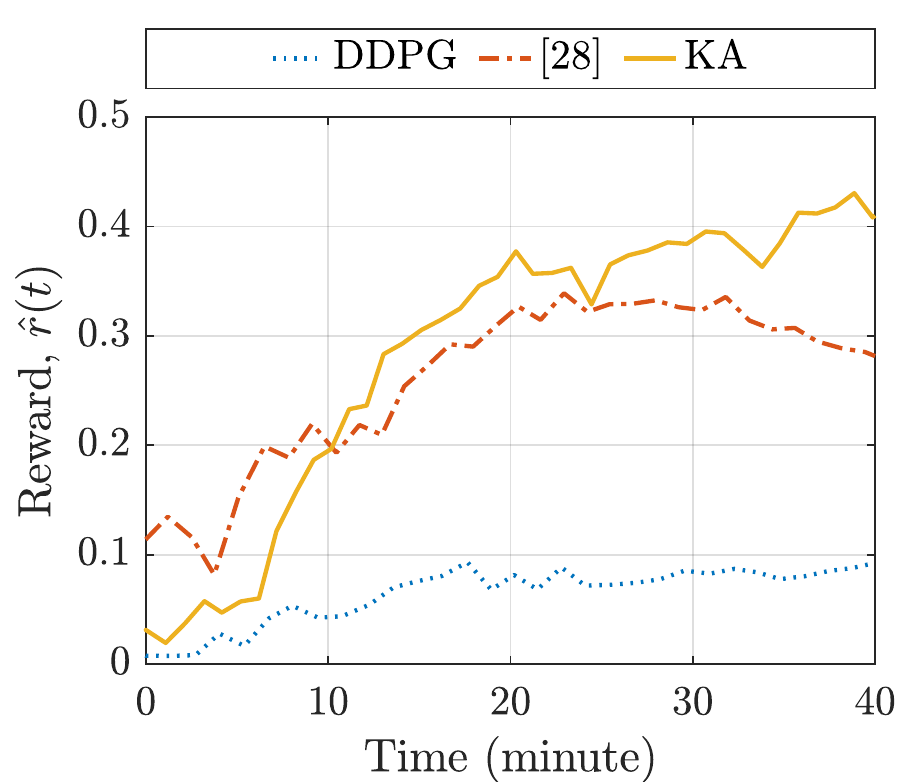}
\caption{Average reward of users.}
\label{subfig:plot_rwd_per_ue_other-avg}
\end{subfigure}
\begin{subfigure}{.8\columnwidth}
\centering
\includegraphics[scale=0.7]{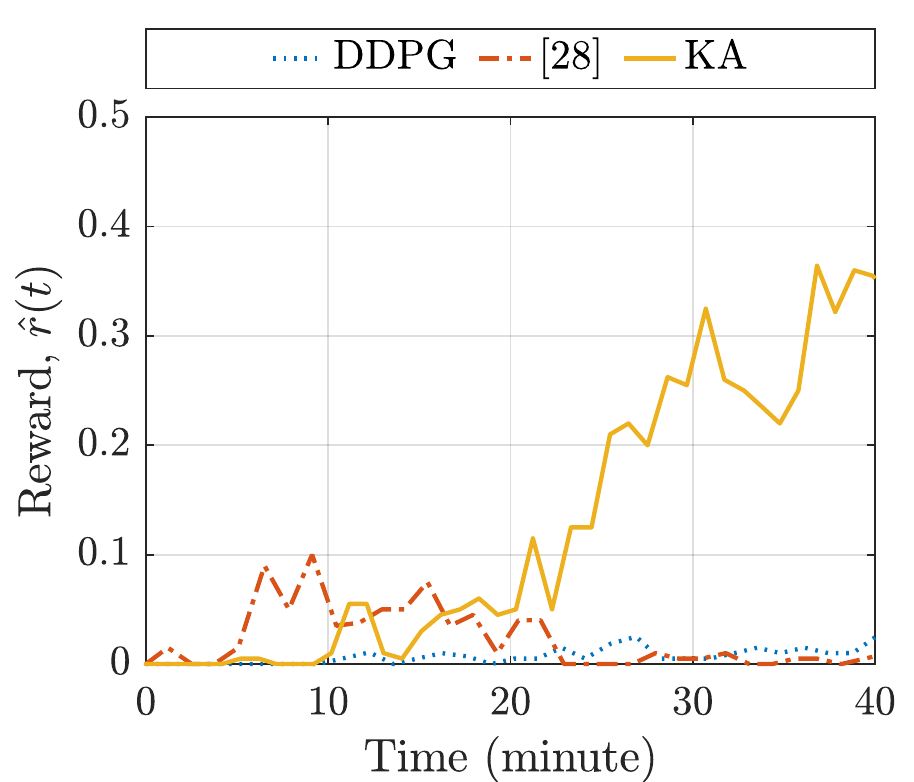}
\caption{Reward of the worst-case user.}
\label{subfig:plot_rwd_per_ue_other-min}
\end{subfigure}
\caption{Rewards of different DDPG in the T-DRL framework, where $K=15$ and $N=50$.}\label{fig:plot_rwd_per_ue_other}
\vspace{-0.2cm}
\end{figure}

\begin{figure}[t]
\centering
\begin{subfigure}{.8\columnwidth}
\centering
\includegraphics[scale=0.7]{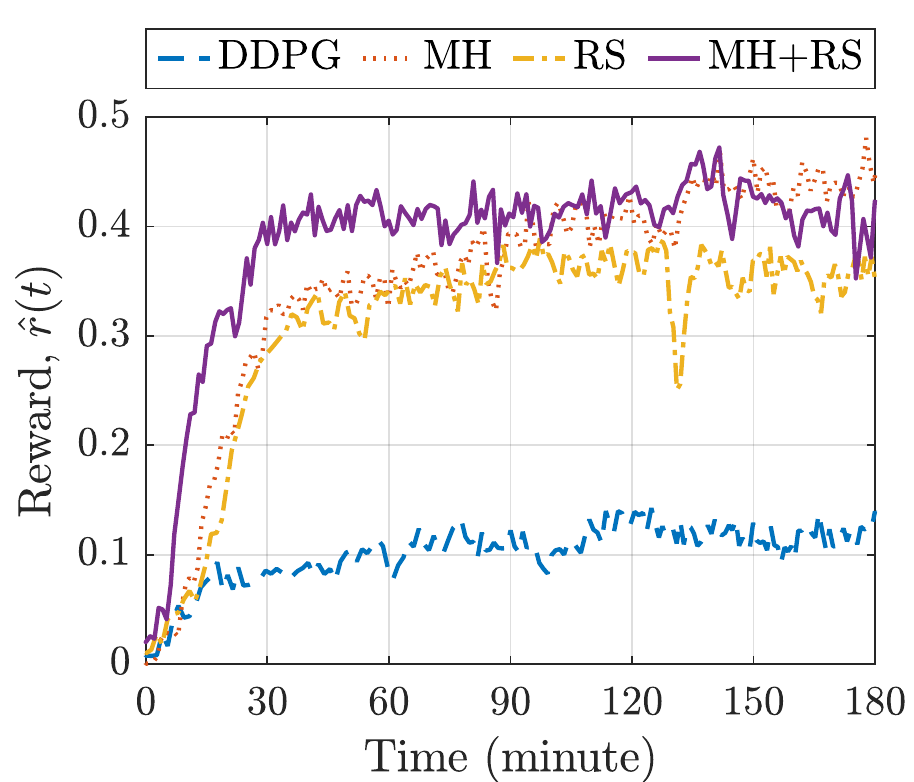}
\caption{Average reward of users.}
\label{subfig:plot_rwd_per_ue_self-avg}
\end{subfigure}
\begin{subfigure}{.8\columnwidth}
\centering
\includegraphics[scale=0.7]{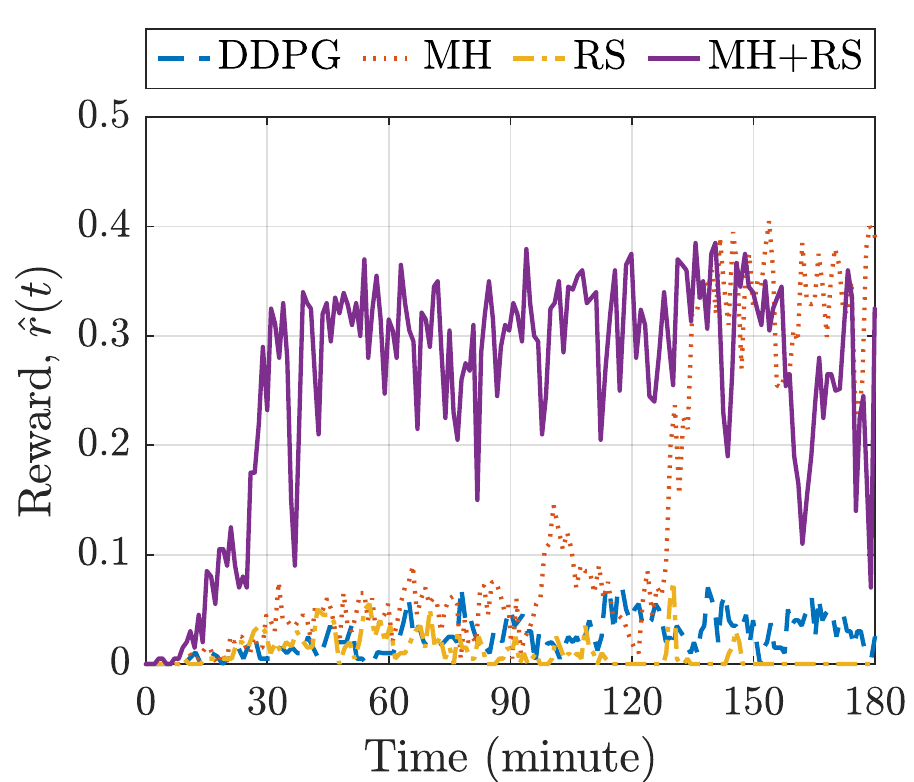}
\caption{Reward of the worst-case user.}
\label{subfig:plot_rwd_per_ue_self-min}
\end{subfigure}
\caption{Rewards of DDPG with the assistance of different kinds of knowledge in the T-DRL framework, where $K=15$ and $N=50$.}\label{fig:plot_rwd_per_ue_self}
\vspace{-0.2cm}
\end{figure}

\begin{figure}[!ht]
\centering
\includegraphics[scale=0.8]{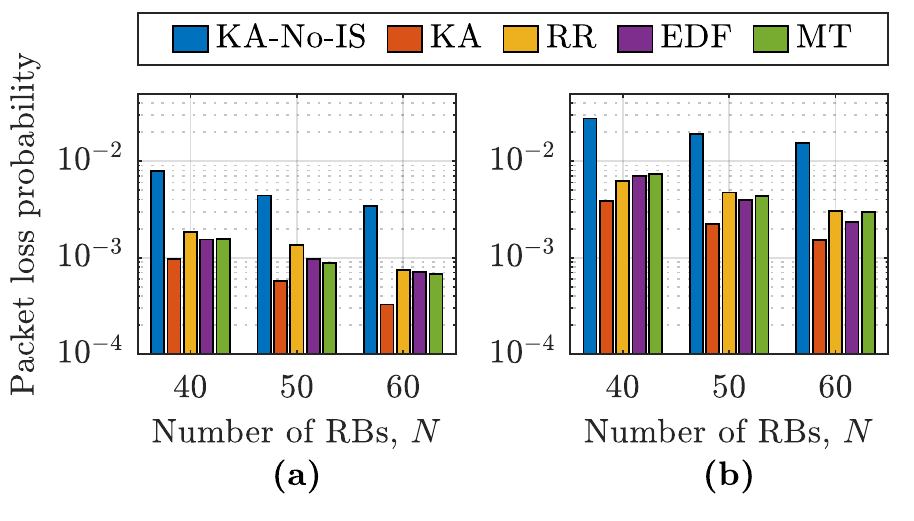}
\caption{Reliability of scheduler for different total numbers of RBs, where $K=15$. (a) Average packet loss probability of all users; (b) Packet loss probability of the worst-case user. }
\label{fig:eval_reliability}
\vspace{-0.2cm}
\end{figure}

We then compare the reward of different DDPG algorithms in the T-DRL framework, including the original DDPG (with legend ``DDPG"), an extension of DDPG in \cite{gu2019intelligent} (with legend ``\cite{gu2019intelligent}"), and our K-DDPG (with legend ``KA"). Fig.~\ref{subfig:plot_rwd_per_ue_other-avg} shows the average reward of users achieved by these three schemes in a 40-minute training phase. The results show that the algorithm in \cite{gu2019intelligent} achieves a higher reward at the beginning of the training phase. This is because a human-written scheduler is used in exploration, which has a better performance than the randomly initialized actor in our scheme. However, our K-DDPG learns faster and achieves better performance than the other two algorithms by the end of the training phase. The reward of the worst-case user is shown in Fig.~\ref{fig:plot_rwd_per_ue_other}. The result indicates that K-DDPG is much better than two other schemes.

To better illustrate the benefits of different kinds of knowledge, we illustrate the reward achieved by different algorithms: 1) original DDPG; 2) DDPG that exploits knowledge of the reward structure by using multi-head critic (with legend ``MH"); 3) DDPG that exploits knowledge of the target scheduling policy by using reward shaping (with legend ``RS"); 4) DDPG with both multi-head critic and reward shaping (with legend ``MH+RS").
The result in Fig.~\ref{fig:plot_rwd_per_ue_self} indicates that multi-head critic helps improve the average reward and the reward of the worst-case user significantly. For example, the average reward of ``MH" in Fig.~\ref{subfig:plot_rwd_per_ue_self-avg} is $3$ times higher than the average reward of DDPG. From the results in Fig.~\ref{subfig:plot_rwd_per_ue_self-min} we can see that if reward shaping is further applied, the convergence time can be reduced by $80$\% from $140$ minutes to $30$ minutes. Note that the reward of ``MH+RS" starts to decrease after 150 minutes of training. This is due to the overfitting of NNs. In practice, we only need to train the actor for 30 minutes with ``MH+RS".

The reliability achieved by different schedulers is shown in Fig.~\ref{fig:eval_reliability}. The packet loss probabilities are evaluated over $2000$ episodes. To show the benefit of importance sampling, we evaluate the reliability of K-DDPG with/without importance sampling (with legends ``KA" and ``KA-No-IS", respectively). In addition, we also evaluate the reliability of three existing schedulers: the round-robin scheduler (with legend ``RR"), the earliest-deadline-first scheduler (with legend ``EDF") and the maximum throughput scheduler (with legend ``MT"). The average packet loss probabilities of all users and the packet loss probabilities of the worst-case user are provided in Fig.~\ref{fig:eval_reliability}a and Fig.~\ref{fig:eval_reliability}b, respectively. The results indicate that without importance sampling, the scheduler can hardly achieve high reliability, while by using importance sampling, K-DDPG can reduce the packet loss probability by $30\% \sim 50$\% compared with the three existing schedulers.

\section{Prototype of Proposed Online Architecture and Experimental Results}\label{sec:experiment}
In this section, we show how to implement the proposed online DDPG architecture in a real-world network. Since 5G NR testbed \cite{nikaein2014openairinterface} is still under development and is not available, we use an open-source Long Term Evolution (LTE) software suite \cite{gomez2016srslte} to build the prototype, in which we measure the processing time in both inference and training as well as the E2E latency and rewards experienced by users.

\subsection{Prototype}
\subsubsection{Proposed architecture}
\begin{figure}[!t]
\centering
\includegraphics[scale=0.75]{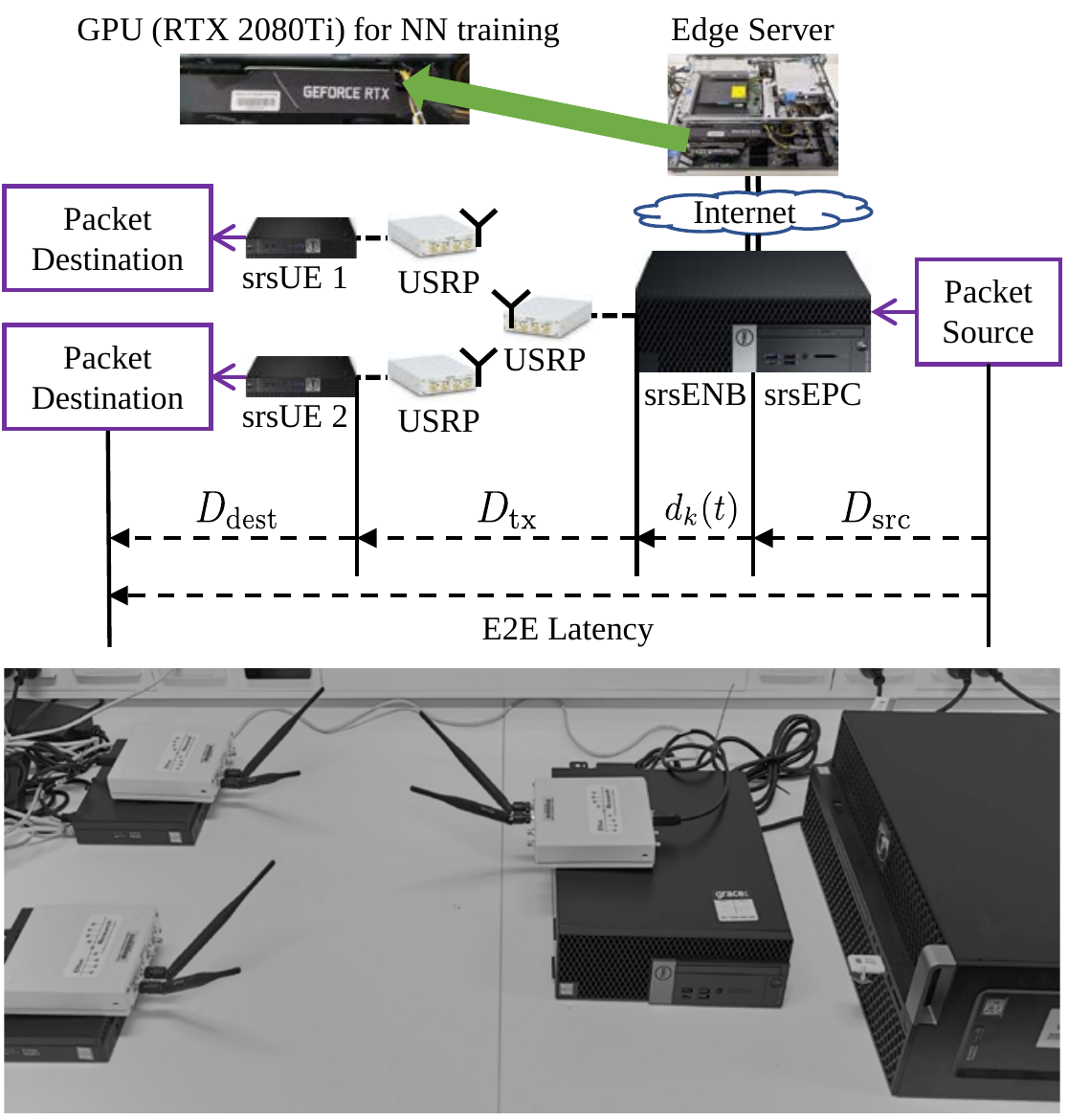}
\caption{A diagram of the experiment setup.}\label{fig:setup_diagram}
\label{fig:exp_setup}
\vspace{-0.2cm}
\end{figure}

The diagram of the prototype is shown in Fig.~\ref{fig:setup_diagram}. We built K-DDPG based on the T-DRL framework by using Pytorch in Python \cite{paszke2019pytorch}.
The algorithms run on a Dell 7820 workstation equipped with an RTX2080Ti graphics processing unit (GPU) and two Intel Xeon Gold 6134 central processing units (CPUs) with 8 cores each. 
The action generation process is developed based on libtorch in C++ \cite{paszke2019pytorch}.
We constructed the standard-compliant cellular network based on the open-source LTE software suite developed by Software Radio System (srsLTE) \cite{gomez2016srslte}, which consists of eNodeB (srsENB, the BS), evolved packet core (srsEPC, the core network) and user equipment (srsUE, the user). 
We embedded the action generation process in the scheduler of srsENB.
srsENB and srsEPC run on a Dell 7060 computer that has an Intel i7-8700 CPU with 6 cores and srsUE run on Dell 7050 computers equipped with an Intel i7-6700 CPU with 4 cores. The radio transceivers for the BS and users are universal software radio peripheral (USRP) B210.
We developed the communication protocol between the server and the BS in Google Protocol Buffers that can automatically compile the protocol into Python and C++. We set the number of RBs, $N$, as 15. The duration of each slot, $\Delta^t$, is 1 ms and the bandwidth of each RB, including 12 subcarriers in LTE, is $W=180$ kHz.

\subsubsection{Measurement platform}
We developed a measurement platform to carry out experiments and measurements. In the platform, the packet source sends user datagram protocol (UDP) packets to the packet destinations.
The packet size is 150 bytes and the arrival rate is $0.1$~packet/ms.
The E2E latency of a packet is measured at the packet destination by comparing the time it is sent by the source and the time it is received by the destination. This requires the clocks of the computers are highly synchronized. 
To achieve this goal, we implemented a clock synchronization system based on precision time protocol, which synchronizes the clocks of the computers at a sub-microsecond level and allows accurate E2E latency measurements.
Note that the clock synchronization system is not required to deploy T-DRL framework and K-DDPG in a commercial cellular network.

\subsection{Mismatch between Simulation and Real-world Network}
In the simulation, the time is discretized into slots (i.e., TTIs).
Thus, the HoL delays are integers. In the real-world BS, the measured HoL delays are with nanosecond precision, denoted by $\hat{d}_k$, for $k=1,\dots,K$.
We convert it to the number of slots from $d_k=\text{round}(\hat{d}_k/\Delta^t)$, where $\text{round}(x)$ is the closest integer to $x$.
Furthermore, the CSI in the real-world BS is reported by users, i.e., a four-bit binary number referred to as the channel quality indicator. We can map this channel quality indicator to SNR based on the method in \cite{3GPP2017Agree,mogensen2007lte}.
E2E latency in the real-world network includes the delay from the packet source to srsENB, $D_\text{src}$, the queueing delay at the srsENB, $d_k(t)$, the transmission delay, $D_\text{tx}$, and the delay from srsUE to the packet destination, $D_\text{dest}$,
We denote the total delays excluding the queueing delay as $D_\text{other} \triangleq D_\text{src} + D_\text{tx} +D_\text{dest}$.
To meet the QoS requirements of time-sensitive traffic, the E2E latency should lie in $[D_\text{min}+D_\text{other},D_\text{max}+D_\text{other}]$.
We assume that  $D_\text{src} \ll D_\text{tx}$ and $D_\text{dest} \ll D_\text{tx}$. Then, $D_\text{other} \approx D_\text{tx} = 4$~ms in the LTE system \cite{3gpp.36.213}.

\subsection{Tests in a Real-world Network}
\begin{figure}[t]
\centering
\begin{subfigure}{1\columnwidth}
\centering
\includegraphics[scale=0.81]{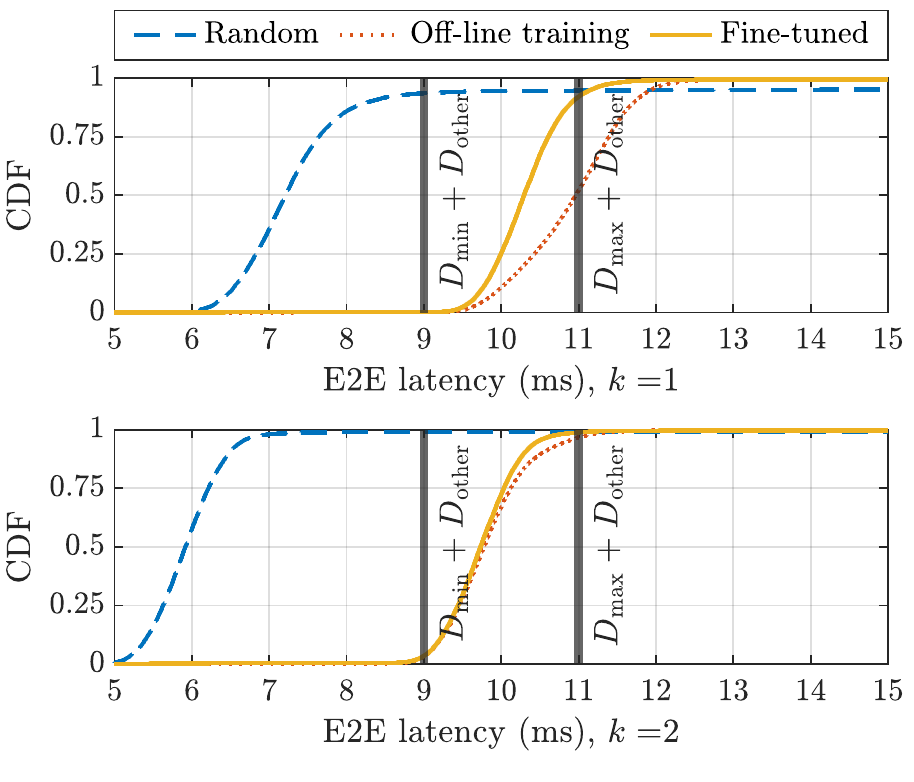}
\caption{E2E latency in the prototype.}\label{fig:plot_online_init}
\end{subfigure}
\\
\begin{subfigure}{1\columnwidth}
\centering
\includegraphics[scale=0.81]{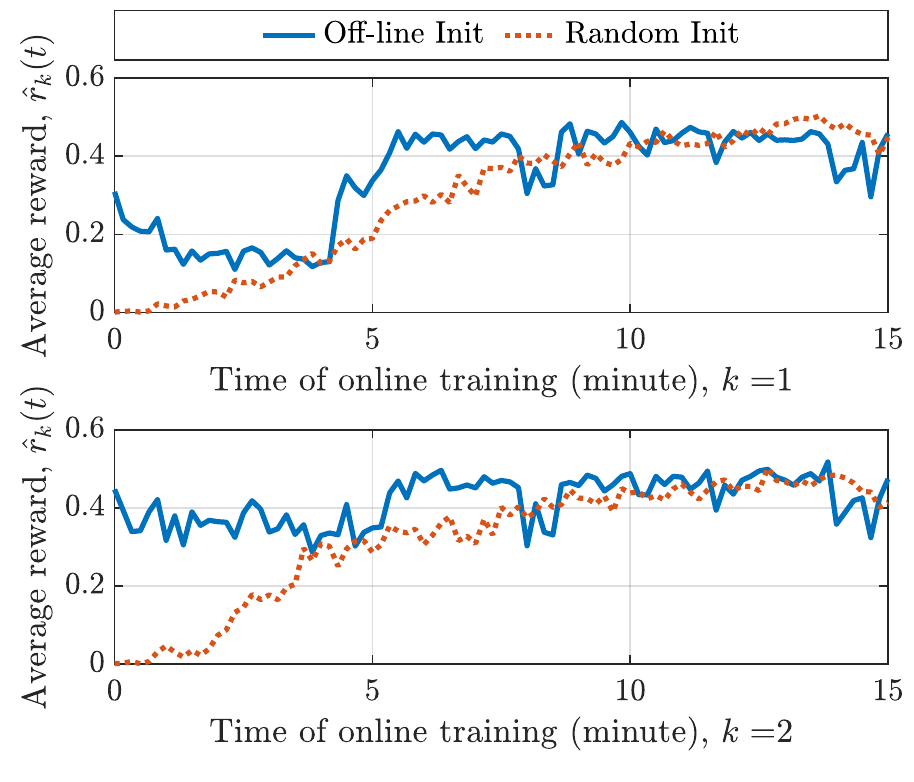}
\caption{Average rewards during online training.}\label{fig:plot_online_rwd}
\end{subfigure}
\caption{The results of online training in the proposed architecture.}\label{fig:plot_online}
\vspace{-0.2cm}
\end{figure}
Fig.~\ref{fig:plot_online_init} compares the cumulative distribution function (CDF) of the E2E latency experienced by two users, which is measured in the prototype for 2 minutes.
The results show that with a high probability, the latency achieved by NNs initialized with random parameters (with legend ``Random") does not lie in $[D_\text{min}+D_\text{other},D_\text{max}+D_\text{other}]$.
For the actor trained off-line in the simulation platform and directly applied in the real-world network without fine-tuning (with legend ``Off-line training"), with high probability $d_2(t) \in [D_\text{min}+D_\text{other},D_\text{max}+D_\text{other}]$, but only around half of the packets are delivered to the first user with $d_1(t) \in [D_\text{min}+D_\text{other},D_\text{max}+D_\text{other}]$. This is because the simulation platform is not exactly the same as the prototype.
To handle this issue, the online DDPG architecture is applied to fine-tune the pre-trained actor and critic in the prototype with a 15-minute online training phase, where we tried different configurations of the parameter space noise for online exploration, $\theta^\mu  \leftarrow \theta^\mu \cdot (1+ \mathbb{N}(0,v^2) \cdot e ^ {-\lambda t \Delta^t})$, that is described in Section \ref{subsec:online_training}. We set the parameters as $v=0.1$ and $\lambda=5\times10^4$, because these values can achieve the best performance according to our experience. With the fine-tuned actor (with legend ``Fine-tuned"), the probability that $d_1(t) \in [D_\text{min}+D_\text{other},D_\text{max}+D_\text{other}]$ is improved remarkably and performance the second user is also improved slightly. The average rewards of two users during online training are shown in Fig.~\ref{fig:plot_online_rwd}. The parameters of the actor and the critic are either initialized off-line in our simulation platform (with legend ``Off-line Init") or initialized with random variables (with legend ``Random Init"). 
It shows that off-line initialization not only significantly improves the initial performance but also reduces the convergence time of users by at least $40$\%.\footnote{By comparing off-line initialization with random initialization, which is the usual case in most of deep learning algorithms, we intended to show how much training time can be saved by off-line initialization in the prototype. Few-shot learning methods can be applied to further reduce the online training time \cite{jadon2020overview}. However, this is beyond the research scope of this paper.} If the environment is highly dynamic (e.g., high mobility, burst traffic pattern, and frequent user list update), We might not be able to adjust the hyper-parameters of the actor and the critic in time, e.g., adjusting the number of hidden layers and the number of neurons in each layer. To handle this issue, one may consider applying few-shot learning methods \cite{jadon2020overview} to further reduce the time needed for online fine-tuning. Also, one can use graph neural networks to transfer the trained NNs into scheduler design problems with different scales~\cite{eisen2020optimal}.

We measured the processing time of the feed-forward inference of the actor that runs on the Intel i7-8700 CPU at the BS.
The average processing time of the inference is $0.036$~ms and the maximum processing time is $0.067$~ms, which is less than the duration of the shortest TTI in 5G NR, e.g., $0.125$~ms. This result indicates that our scheduler can be operated at every TTI in real-world 5G systems. We also observed that the processing time grows as the sizes of the NNs increase. When the sizes of the NNs are large, we may need GPUs, field-programmable gate arrays or application-specific integrated circuits at the BS in order to avoid processing delay violation.
Furthermore, we measured the average processing time of each training iteration in the edge server, i.e., around $5$~ms. Thus, the online DDPG architecture can update the actor according to real-world networks every few milliseconds.




\section{Conclusion}\label{sec:conclusion}
In this paper, we implement K-DDPG in wireless scheduler design for time-sensitive traffic in 5G NR. We found that the straightforward implementation of DDPG converges slowly, has a poor QoS performance, and can hardly be implemented in real-world 5G NR systems. To address these issues, we first proposed a T-DRL framework based on the theoretical models and results. Then, different kinds of expert knowledge of the scheduler design problem were exploited to reduce the convergence time and to improve the individual QoS of each user. Furthermore, we developed an online DDPG architecture that enables off-line initialization and online fine-tuning.
Our simulation and experimental results indicated that by using K-DDPG in the T-DRL framework, the convergence time and the individual QoS of each user can be improved significantly. In addition, with our online architecture, the scheduling policy can be updated according to real-world feedback every few milliseconds, and can be executed in each TTI in 5G NR.


\section*{Appendix: Proof of the Markov Property}\label{sec:mdp}
To apply DRL, we prove the Markov property in this subsection. We first derive the transition probability of HoL delay. 
\begin{figure}[!t]
        \centering
        \includegraphics[scale=0.8]{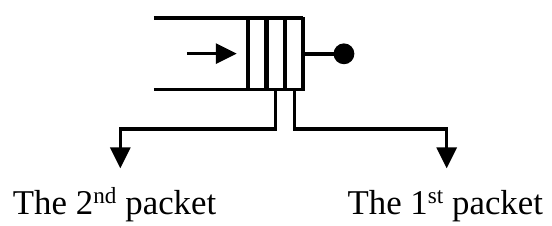}
        \caption{Illustration of queueing model.}
        \label{fig:hol_delay}
        \vspace{-0.5cm}
\end{figure}

If the $k$-th user is scheduled in the $t$-th slot, the transition probability is denoted by $p_{i,j}^{k,+} = \Pr\{d_k(t+1) = j\ |\ d_k(t) = i, x_k(t) = 1\}$, where $i$ is the HoL delay in the $t$-th slot and $j$ is the HoL delay in the $t+1$-th slot. Since users with empty buffers will not be scheduled, we have $i > 0$. To derive the transition probability, we consider the following three cases: 1) $0 < i < j$, 2) $0< j \leq i$ and 3) $0 = j < i$. As shown in Fig.~\ref{fig:hol_delay}, the queueing delays of the first and the second packets in the $t$-th slot are $i$ and $j-1$, respectively. The inter-arrival time between the first and the second packet is $i-(j-1)$. Since the inter-arrival time is strictly positive, we have $i-(j-1) > 0 $ and $j \leq i$. This means that for all $0 < i < j$, $p_{i,j}^{k,+}=0$. For the case $0< j \leq i$, $p_{i,j}^{k,+}$ equals to the probability that the second packet arrived at the buffer $i-(j-1)$ slots later than the first packet. For the Bernoulli arrival process, $p_{i,j}^{k,+} = p_k(1 - p_k)^{i-j}$. For the case $j=0$, the buffer becomes empty in the $(t+1)$-th slot. It means that no packet arrived at the buffer during the past $i$ slots. Thus, $p_{i,j}^{k,+} = (1 - p_k)^{i}$.

If the $k$-th user is not scheduled in the $t$-th slot, the transition probability is denoted by $p_{i,j}^{k,-} = \Pr\{d_k(t+1) = j\ |\ d_k(t) = i, x_k(t) = 0\}$. To derive $p_{i,j}^{k,-}$, we consider three cases: 1) the buffer is empty, $i=0$, 2)  $0< i < D_{\max}$ and 3) $i = D_{\max}$.
When the buffer is empty in the $t$-th slot, $i=0$. With probability $p_k$, a packet arrives at the buffer in the $t$-th slot and $j=1$. Otherwise, $j=0$. When the HoL delay is smaller than the maximum delay bound, $0< i <D_\text{max}$, the HoL delay will increase by one slot. When the HoL delay equals the maximum delay bound, $i=D_\text{max}$, the first packet will be discarded. The HoL delay in the next slot depending on the queueing delay of the second packet. If the second packet arrived within the previous $D_{\max}$ slots, $p_{i,j}^{k,-} = p_k (1 - p_k )^{D_\text{max}-j}$. Otherwise, $p_{i,j}^{k,-} = (1 - p_k )^{D_\text{max}}$.

Since the above transition probabilities only depend on the states and actions in the $t$-th slot, the HoL delay is Markovian. By assuming that the wireless channel fading is Markovian, the problem is an optimal control problem of a Markov decision process. This completes the proof.

\bibliography{main}

\begin{thebibliography}{10}
\providecommand{\url}[1]{#1}
\csname url@samestyle\endcsname
\providecommand{\newblock}{\relax}
\providecommand{\bibinfo}[2]{#2}
\providecommand{\BIBentrySTDinterwordspacing}{\spaceskip=0pt\relax}
\providecommand{\BIBentryALTinterwordstretchfactor}{4}
\providecommand{\BIBentryALTinterwordspacing}{\spaceskip=\fontdimen2\font plus
\BIBentryALTinterwordstretchfactor\fontdimen3\font minus
  \fontdimen4\font\relax}
\providecommand{\BIBforeignlanguage}[2]{{%
\expandafter\ifx\csname l@#1\endcsname\relax
\typeout{** WARNING: IEEEtran.bst: No hyphenation pattern has been}%
\typeout{** loaded for the language `#1'. Using the pattern for}%
\typeout{** the default language instead.}%
\else
\language=\csname l@#1\endcsname
\fi
#2}}
\providecommand{\BIBdecl}{\relax}
\BIBdecl

\bibitem{aijaz2018tactile}
A.~Aijaz and M.~Sooriyabandara, ``The tactile internet for industries: a
  review,'' \emph{Proc. IEEE}, vol. 107, no.~2, pp. 414--435, 2018.

\bibitem{bennis2018ultrareliable}
M.~Bennis, M.~Debbah, and H.~V. Poor, ``Ultrareliable and low-latency wireless
  communication: Tail, risk, and scale,'' \emph{Proc. IEEE}, vol. 106, no.~10,
  pp. 1834--1853, 2018.

\bibitem{she2017radio}
C.~She, C.~Yang, and T.~Q. Quek, ``Radio resource management for ultra-reliable
  and low-latency communications,'' \emph{IEEE Commun. Mag.}, vol.~55, no.~6,
  pp. 72--78, 2017.

\bibitem{3gpp.22.104}
3GPP, ``{Service requirements for cyber-physical control applications in
  vertical domains},'' {3GPP}, TS 22.104, 2018, v16.0.0.

\bibitem{huang2018gpf}
Y.~Huang, S.~Li, Y.~T. Hou, and W.~Lou, ``{GPF}: {A} {GPU}-based design to
  achieve\~{} 100 $\mu$s scheduling for 5{G} {NR},'' in \emph{MobiCom}, 2018.

\bibitem{kawser2012performance}
M.~T. Kawser, H.~Farid, A.~R. Hasin, A.~M. Sadik, and I.~K. Razu, ``Performance
  comparison between round robin and proportional fair scheduling methods for
  lte,'' \emph{International Journal of Information and Electronics
  Engineering}, vol.~2, no.~5, pp. 678--681, 2012.

\bibitem{andrews2000probabilistic}
M.~Andrews, ``Probabilistic end-to-end delay bounds for earliest deadline first
  scheduling,'' in \emph{IEEE INFOCOM}, 2000.

\bibitem{schwarz2010low}
S.~Schwarz, C.~Mehlf{\"u}hrer, and M.~Rupp, ``Low complexity approximate
  maximum throughput scheduling for {LTE},'' in \emph{IEEE ASILOMAR}, 2010.

\bibitem{she2020deep}
C.~She, R.~Dong, Z.~Gu \emph{et~al.}, ``Deep learning for ultra-reliable and
  low-latency communications in 6{G} networks,'' \emph{IEEE Network, accepted},
  2020.

\bibitem{mnih2015human}
V.~Mnih, K.~Kavukcuoglu, D.~Silver \emph{et~al.}, ``Human-level control through
  deep reinforcement learning,'' \emph{Nature}, vol. 518, no. 7540, p. 529,
  2015.

\bibitem{sutton2011reinforcement}
R.~S. Sutton and A.~G. Barto, \emph{Reinforcement learning: An
  introduction}.\hskip 1em plus 0.5em minus 0.4em\relax Cambridge, MA: MIT
  Press, 2011.

\bibitem{lillicrap2015continuous}
T.~P. Lillicrap, J.~J. Hunt, A.~Pritzel \emph{et~al.}, ``Continuous control
  with deep reinforcement learning,'' \emph{arXiv preprint arXiv:1509.02971},
  2015.

\bibitem{nasrallah2018ultra}
A.~Nasrallah, A.~S. Thyagaturu, Z.~Alharbi \emph{et~al.}, ``Ultra-low latency
  ({ULL}) networks: The {IEEE} {TSN} and {IETF} {DetNet} standards and related
  {5G} {ULL} research,'' \emph{IEEE Commun. Surveys Tuts.}, vol.~21, no.~1, pp.
  88--145, 2018.

\bibitem{specht2017synthesis}
J.~Specht and S.~Samii, ``Synthesis of queue and priority assignment for
  asynchronous traffic shaping in switched ethernet,'' in \emph{RTSS}, 2017.

\bibitem{khoshnevisan20195g}
M.~Khoshnevisan, V.~Joseph, P.~Gupta \emph{et~al.}, ``5{G} industrial networks
  with {CoMP} for u{RLLC} and time sensitive network architecture,'' \emph{IEEE
  J. Sel. Areas Commun.}, vol.~37, no.~4, pp. 947--959, 2019.

\bibitem{ginthor2019analysis}
D.~Ginth{\"o}r, J.~von Hoyningen-Huene, R.~Guillaume, and H.~Schotten,
  ``Analysis of multi-user scheduling in a {TSN}-enabled 5{G} system for
  industrial applications,'' in \emph{IEEE ICII}, 2019.

\bibitem{3GPP2012MTC}
3GPP, ``Analysis on traffic model and characteristics for {MTC} and text
  proposal,'' {3GPP}, TR R1-120056, 2012, {TSG-RAN} Meeting WG1\#68, Dresden,
  Germany.

\bibitem{Hassan2013A}
H.~A. Omar, W.~Zhuang, A.~Abdrabou, and L.~Li, ``A feasibility study and
  development framework design for realizing smartphone-based vehicular
  networking systems,'' \emph{IEEE Trans. Emerg. Topics Comput.}, vol.~1,
  no.~1, pp. 69 -- 83, Aug. 2013.

\bibitem{3GPP2017Scenarios}
3GPP, ``Study on scenarios and requirements for next generation access
  technologies,'' {3GPP}, TR 38.913, 2017, v14.2.0.

\bibitem{tseng2019radio}
S.-C. Tseng, Z.-W. Liu, Y.-C. Chou, and C.-W. Huang, ``Radio resource
  scheduling for 5{G} {NR} via deep deterministic policy gradient,'' in
  \emph{IEEE ICC Workshops}, 2019.

\bibitem{qi2019deep}
C.~Qi, Y.~Hua, R.~Li \emph{et~al.}, ``Deep reinforcement learning with discrete
  normalized advantage functions for resource management in network slicing,''
  \emph{IEEE Commun. Lett.}, vol.~23, no.~8, pp. 1337--1341, 2019.

\bibitem{li2020deep}
J.~Li and X.~Zhang, ``Deep reinforcement learning based joint scheduling of
  e{MBB} and u{RLLC} in 5{G} networks,'' \emph{IEEE Commun. Lett.}, 2020.

\bibitem{ayala2019vrain}
J.~A. Ayala-Romero, A.~Garcia-Saavedra, M.~Gramaglia \emph{et~al.}, ``vr{AI}n:
  A deep learning approach tailoring computing and radio resources in
  virtualized {RAN}s,'' in \emph{MobiCom}, 2019.

\bibitem{foukas2019iris}
X.~Foukas, M.~K. Marina, and K.~Kontovasilis, ``Iris: Deep reinforcement
  learning driven shared spectrum access architecture for indoor neutral-host
  small cells,'' \emph{IEEE J. Sel. Areas Commun.}, vol.~37, no.~8, pp.
  1820--1837, 2019.

\bibitem{ng1999policy}
A.~Y. Ng, D.~Harada, and S.~Russell, ``Policy invariance under reward
  transformations: Theory and application to reward shaping,'' in \emph{ICML},
  1999.

\bibitem{he2019model}
H.~He, S.~Jin, C.-K. Wen \emph{et~al.}, ``Model-driven deep learning for
  physical layer communications,'' \emph{IEEE Wireless Commun.}, vol.~26,
  no.~5, pp. 77--83, 2019.

\bibitem{he2020model}
H.~He, C.-K. Wen, S.~Jin, and G.~Y. Li, ``Model-driven deep learning for {MIMO}
  detection,'' \emph{IEEE Trans. Signal Process.}, vol.~68, pp. 1702--1715,
  2020.

\bibitem{gu2019intelligent}
L.~Gu, D.~Zeng, W.~Li \emph{et~al.}, ``Intelligent {VNF} orchestration and flow
  scheduling via model-assisted deep reinforcement learning,'' \emph{IEEE J.
  Sel. Areas Commun.}, vol.~38, no.~2, pp. 279--291, 2019.

\bibitem{gomez2016srslte}
I.~Gomez-Miguelez, A.~Garcia-Saavedra, P.~D. Sutton \emph{et~al.}, ``srs{LTE}:
  an open-source platform for {LTE} evolution and experimentation,'' in
  \emph{WiNTECH}, 2016.

\bibitem{neumann2018towards}
A.~Neumann, L.~Wisniewski, R.~S. Ganesan \emph{et~al.}, ``Towards integration
  of industrial ethernet with {5G} mobile networks,'' in \emph{IEEE WFCS},
  2018.

\bibitem{3gpp.38.214}
3GPP, ``{Physical layer procedures for data},'' {3GPP}, TS 38.214, 2018,
  v15.2.0.

\bibitem{Yury2014Quasi}
W.~Yang, G.~Durisi, T.~Koch, and Y.~Polyanskiy, ``Quasi-static multiple-antenna
  fading channels at finite blocklength,'' \emph{IEEE Trans. Inf. Theory},
  vol.~60, no.~7, pp. 4232--4264, Jul. 2014.

\bibitem{She2018Joint}
C.~She, C.~Yang, and T.~Q.~S. Quek, ``Joint uplink and downlink resource
  configuration for ultra-reliable and low-latency communications,'' \emph{IEEE
  Trans. Commun.}, vol.~66, no.~5, pp. 2266--2280, May 2018.

\bibitem{WirelessCom}
A.~Goldsmith, \emph{Wireless Communications}.\hskip 1em plus 0.5em minus
  0.4em\relax Cambridge University Press, 2005.

\bibitem{van2017hybrid}
H.~Van~Seijen, M.~Fatemi, J.~Romoff \emph{et~al.}, ``Hybrid reward architecture
  for reinforcement learning,'' in \emph{NIPS}, 2017.

\bibitem{schaul2015prioritized}
T.~Schaul, J.~Quan, I.~Antonoglou, and D.~Silver, ``Prioritized experience
  replay,'' in \emph{ICLR}, 2015.

\bibitem{plappert2018parameter}
M.~Plappert, R.~Houthooft, P.~Dhariwal \emph{et~al.}, ``Parameter space noise
  for exploration,'' in \emph{ICLR}, 2018.

\bibitem{3GPP2017Agree}
3GPP, ``Study on {N}ew {R}adio ({NR}) access technology; physical layer aspects
  ({R}elease 14),'' {3GPP}, TR 38.802, 2017, v2.0.0.

\bibitem{nikaein2014openairinterface}
N.~Nikaein, M.~K. Marina, S.~Manickam \emph{et~al.}, ``{OpenAirInterface}: {A}
  flexible platform for {5G} research,'' \emph{ACM SIGCOMM CCR}, vol.~44,
  no.~5, pp. 33--38, 2014.

\bibitem{paszke2019pytorch}
A.~Paszke, S.~Gross, F.~Massa \emph{et~al.}, ``Pytorch: An imperative style,
  high-performance deep learning library,'' in \emph{NIPS}, 2019.

\bibitem{mogensen2007lte}
P.~Mogensen, W.~Na, I.~Z. Kov{\'a}cs \emph{et~al.}, ``{LTE} capacity compared
  to the {S}hannon bound,'' in \emph{IEEE VTC Spring}, 2007.

\bibitem{3gpp.36.213}
3GPP, ``{Physical layer procedures},'' {3GPP}, TS 36.213, 2009, v8.8.0.

\bibitem{jadon2020overview}
S.~Jadon, ``An overview of deep learning architectures in few-shot learning
  domain,'' \emph{arXiv preprint arXiv:2008.06365}, 2020.

\bibitem{eisen2020optimal}
M.~Eisen and A.~R. Ribeiro, ``Optimal wireless resource allocation with random
  edge graph neural networks,'' \emph{IEEE Trans. Signal Process.}, vol.~68,
  pp. 2977--2991, 2020.

\end{thebibliography}
\bibliographystyle{IEEEtran}

\begin{IEEEbiography}
[{\includegraphics[width=1in,height=1.25in,clip,keepaspectratio]{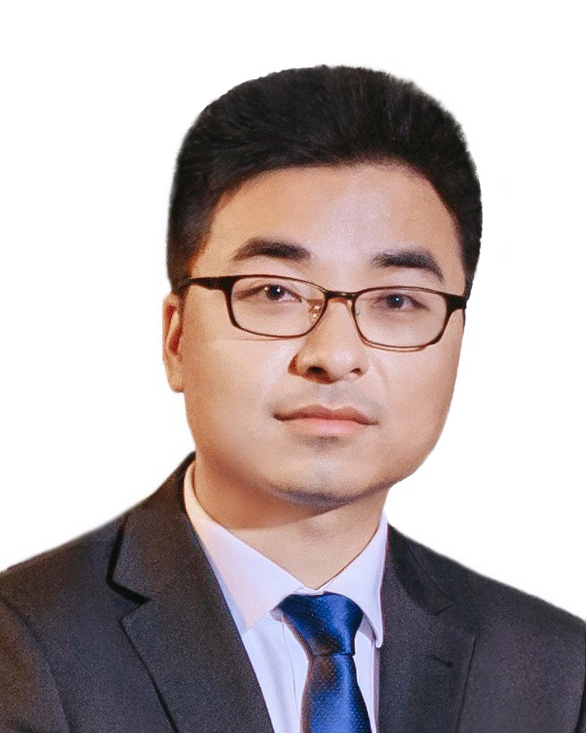}}]{Zhouyou Gu}
received his B.E. degree with First Class Honours and received
his M.Phil. degree from the University of Sydney, Australia, in 2016 and in
2019, respectively. He is currently pursuing his Ph.D. degree in School of Electrical and
Information Engineering at the University of Sydney, Australia. His research
interests focus on the areas of programmable wireless networks, design of radio resource schedulers, and the applications of deep reinforcement learning in 5G and beyond.
\end{IEEEbiography}
\begin{IEEEbiography}
[{\includegraphics[width=1in,height=1.25in,clip,keepaspectratio]{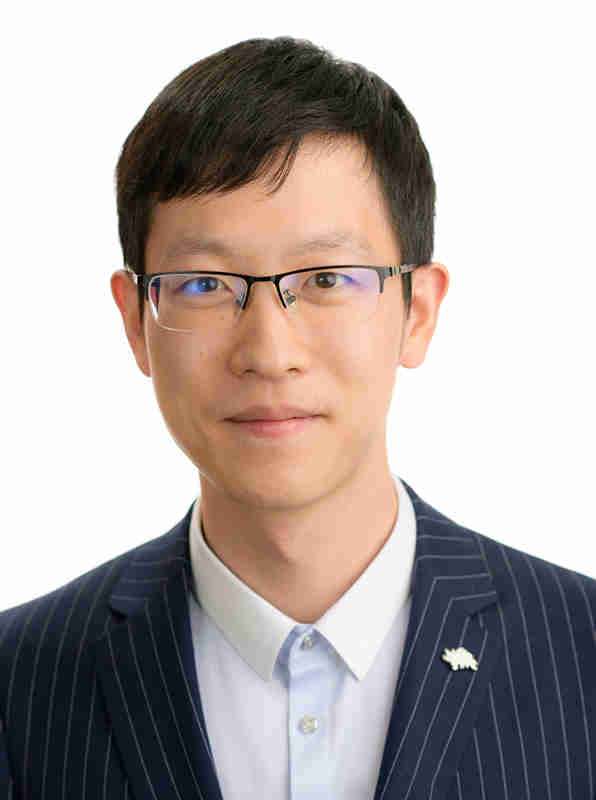}}]{Changyang She}
(S'12-M'17) received his B. Eng degree in Honors College (formerly School of Advanced Engineering) of Beihang University (BUAA), Beijing, China in 2012 and Ph.D. degree in School of Electronics and Information Engineering of BUAA in 2017. From 2017 to 2018, he was a postdoctoral research fellow in Singapore University of Technology and Design. Since 2018, he has become a postdoctoral research associate in the University of Sydney. He is the recipient of the Australian Research Council Discovery Early Career Research Award. His research interests lie in the areas of ultra-reliable and low-latency communications, deep learning in wireless networks, mobile edge computing, and energy efficient 5G communication systems.
\end{IEEEbiography}

\begin{IEEEbiography}
[{\includegraphics[width=1in,height=1.25in,clip,keepaspectratio]{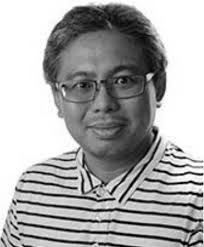}}]{Wibowo Hardjawana}
(M’09) received the Ph.D. degree in electrical engineering from The University of Sydney, Australia, in 2009. He was an Australian Research Council Discovery Early Career Research Award Fellow and is now Senior Lecturer with the School of Electrical and Information Engineering, The University of Sydney. Prior to that he was Assistant Manager at Singapore Telecom Ltd, managing core and radio access networks. His current research interests are in 5/6G cellular radio access and wireless local area networks, with focuses in system architectures, resource scheduling, interference, signal processing and the development of corresponding standard-compliant prototypes.
\end{IEEEbiography}

\begin{IEEEbiography}
[{\includegraphics[width=1in,height=1.25in,clip,keepaspectratio]{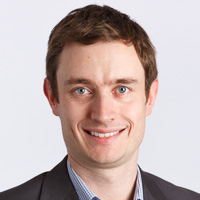}}]{Simon Lumb}
has 15 years’ international wireless telecommunications experience working on 3G, 4G and 5G technologies in operations, network design and configuration as well as research and development. He is currently a Strategic Technology Expert at Telstra and represents the company on the O-RAN Alliance Technical Steering Committee alongside developing Telstra’s future wireless technology strategy. Simon has a B.Eng (Computer Systems Engineering) and B.App.Sci (Computer Science) from RMIT University, Australia.
\end{IEEEbiography}

\begin{IEEEbiography}
[{\includegraphics[width=1in,height=1.25in,clip,keepaspectratio]{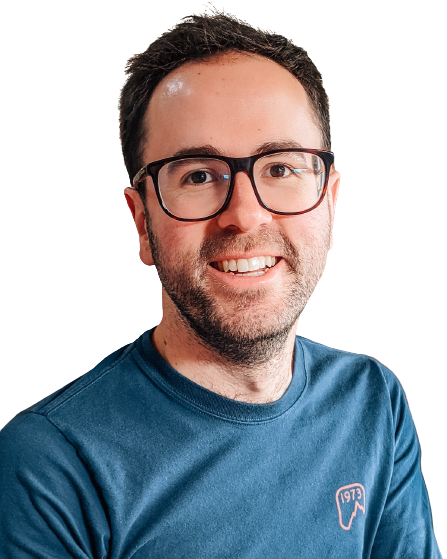}}]{David McKechnie}
is Technology Leader (Future Radio Technologies) at Telstra Corporation, Australia. He is responsible for Telstra's vision and future strategy in wireless networks and connected vehicles. He previously led efforts in low power, wide area networking and Internet of Things/Machine-to-Machine connectivity, and before that, he worked as a senior 4G network engineer deploying mobile networks all over the world. McKechnie has a B. Eng (Electrical and Computer Systems Engineering), from Monash University, Australia.
\end{IEEEbiography}

\begin{IEEEbiography}
[{\includegraphics[width=1in,height=1.25in,clip,keepaspectratio]{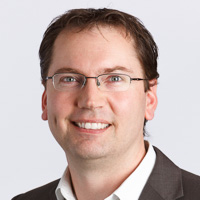}}]{Todd Essery}
(Member, IEEE) has 20 years’ experience in the wireless telecommunications industry, beginning his career at Telstra Research Laboratories (TRL). Early on he executed leading work on wireless machine-to-machine technologies which has since exploded into Internet-of-Things (IoT). Recently as Wireless Technology Area Lead in Telstra Labs, Todd and his team has been developing and executing IoT, 5G, cooperative transport and location technology trials. Todd has a B.Eng (Elec) from the University of Queensland and a MBA from Melbourne Business School (University of Melbourne).
\end{IEEEbiography}

\begin{IEEEbiography}
[{\includegraphics[width=1in,height=1.25in,clip,keepaspectratio]{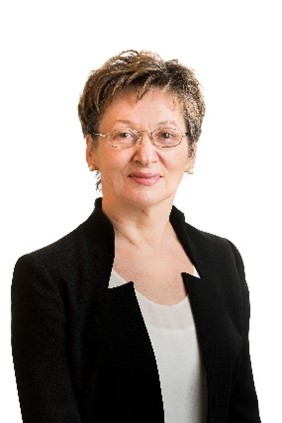}}]{Branka Vucetic} is an ARC Laureate Fellow and Director of the Centre of Excellence for IoT and Telecommunications at the University of Sydney. Her current research work is in wireless networks and the Internet of Things. In the area of wireless networks, she works on ultra-reliable low-latency communications (URLLC) and system design for millimetre wave frequency bands. In the area of the Internet of Things, Vucetic works on providing wireless connectivity for mission critical applications.

Branka Vucetic is a Fellow of IEEE, the Australian Academy of Technological Sciences and Engineering and the Australian Academy of Science. 

\end{IEEEbiography}

\end{document}